\documentclass{mn_plus_bib}
\usepackage{epsfig}
\begin{document}
\title{Towards simulating star formation in the interstellar medium}
\author[Slyz A. et al.]
       {Adrianne D. Slyz$^{1}$, Julien E. G. Devriendt$^{1,2}$, 
        Greg Bryan$^{1}$ and Joseph Silk$^{1}$\\        
       {$^1$Astrophysics, Keble Road, Oxford OX1 3RH, UK}\\
       {$^2$CRAL-Observatoire, 9 Avenue
                Charles-Andr\'{e}, 69561 Saint-Genis-Laval, France}}
\maketitle
\begin{abstract}
As a first step to a more complete understanding of the local physical
processes which determine star formation rates (SFRs) 
in the interstellar medium (ISM), we have performed controlled numerical
experiments consisting of 
hydrodynamical simulations of a kilo-parsec scale, periodic, highly supersonic and
"turbulent" three-dimensional flow. 
Using simple but physically
motivated recipes for identifying star forming regions, we convert gas
into stars which we follow self-consistently as they impact their
surroundings through supernovae and stellar winds. We investigate how
various processes (turbulence, radiative cooling, self-gravity, and
supernovae feedback) structure the ISM, determine its energetics,
and consequently affect its SFR.  We find that
the one-point statistical measurement captured by the probability density
function (PDF) is sensitive to the simulated physics. 
The PDF is consistent with a log-normal distribution for the runs which remove
gas for star formation and have radiative cooling, but implement neither
supernovae feedback nor self-gravity. In this case, the dispersion,
$\sigma_{\mathrm{s}}$, of the log-normal decays with time and 
scales with $\sqrt{{\mathrm {ln}}(1 + ({\mathrm{M}}_{{\mathrm {rms}}}/2)^2)}$ where ${\mathrm{M}}_{\mathrm{rms}}$ is the root-mean-squared
Mach number of the simulation volume, ${\mathrm{s} = {\mathrm{ln}}\; \rho}$ , and $\rho$ is
the gas density. With the addition of 
self-gravity, the log-normal consistently under-predicts the high density end
of the PDF which approaches a power law. With 
supernovae feedback, regardless of whether we consider self-gravity 
or not, the PDF becomes markedly 
bimodal with most of the simulation volume occupied by low density gas.
Aside from its effect on the density structure of the medium, including 
self-gravity and/or supernovae feedback changes the dynamics of the medium
by halting the decay of the kinetic energy.
Since we find that the SFR depends most strongly on the underlying 
velocity field, the SFR declines in the runs lacking a means to sustain the
kinetic energy, and the subsequent high density constrasts. This strong
dependence on the gas velocity dispersion is in
agreement with Silk's formula for the SFR (Silk 2001) which also takes the hot gas porosity,
and the average gas density as important parameters.
Measuring the porosity of the hot gas for the runs with supernovae
feedback, we compare Silk's model for the SFR
to our measured SFR and find agreement to better than a factor two.

\end{abstract}


\section{Introduction}
\label{introduction}
The correlation in galaxies between the star formation rate 
and the average gas surface density over several orders of magnitude 
(Kennicutt 1998) suggests a simple, 
deterministic prescription (Schmidt 1959) for star formation.  
Yet the finding that, at least in the Milky Way,  
all star formation occurs in dense, cold clouds of 
molecular hydrogen and dust raises the question of how information 
about the average gas density of a galaxy reaches the small scale 
on which star formation occurs.
Furthermore, observations of our own interstellar
medium (ISM) as well as that of other galaxies reveal that far from being
well described by a global quantity like the average gas density, the ISM
has a spectacularly complex structure on many scales.
Diffuse ionized gas in edge-on spirals is concentrated in webs of 
filaments and shells (Rand, Kulkarni \& Hester 1990, Dettmar 1992, 
Ferguson, Wyse, \& Gallagher 1996).  Atomic gas detected by 21 cm
emission in our Galaxy (Heiles 1979, 1984) as well as in several 
other spirals (Irwin 1994, Rand \& van der Hulst 1993, Lee \& Irwin 1997, 
King \& Irwin 1997) resides in ``supershells'' and ``worms''. 
In maps of the nearby spirals M31 and M33 (Brinks \& Bajaja 1986, 
Deul \& den Hartog 1990), it is 
also found to be depleted in numerous 100 pc -- 1 kpc ``holes''. 
Attempts to quantify this elaborate
ISM structure are confronted with questions of identification.
Structures are interconnected, with, for example, denser regions of gas 
embedded within filaments. Hence for example, potential sites of star formation
cannot be picked out, without introducing a density threshold and thereby 
a bias to separate them from the underlying density field. 
An alternative way to analyze the ISM is with Fourier transform power spectra.
Applied to HI emission maps of the Large and Small Magellanic Clouds, 
power laws over $\sim$ 2 orders of magnitude are found
(Stanimirovic et al. 1999, Elmegreen, Kim, Stavely-Smith 2001),
providing another insight into the structure of the ISM, namely
that as other observations have already suggested, it is likely to be
turbulent.

Clues about the energy sources for the stirring of the ISM
come from measurements of the sizes and velocities of
shells. In some cases stellar winds and supernovae are found 
to be adequate for creating the supershells, and HI holes. In other
cases larger quantities of energy are demanded and then collisions of 
external clouds with the galaxies are invoked (Tenorio-Tagle 1981).
As for the diffuse ionized medium, although the energy available from
O stars would be sufficient to account for its photoionization,
a well-known problem is that photons from the O stars cannot travel 
far from their origin without being absorbed by the molecular clouds 
and HI halos surrounding them.
In that case the photons either reach larger distances by traveling
through photoionized conduits carved out by earlier supernovae or as suggested
by an alternative model they are additionally generated in 
turbulent mixing layers at the interfaces
between hot and cold gas.  These are ubiquitous in the ISM, and have been 
invoked as an efficient means to 
convert the thermal energy generated by shear flows to ionizing radiation
(Begelman \& Fabian 1990, Slavin, Shull, \& Begelman 1993). 
Ultimately the energy source in the latter model
is again the supernovae which create the hot gas.  Recent X-ray images
from Chandra map out this hot, tenuous gas, predicted 
by Spitzer (1956), above and below the galactic plane of 
disk galaxies (Wang et al. 2001). 
Even without a heat source due to its long cooling time, 
once it is generated by supernovae, such gas can persist for 
millions of years. Cox \& Smith (1974) reasoned that
given that OB stars occur in associations, it is likely that
a supernovae will go off inside the hot cavity generated by a previous
supernovae, thereby rejuvenating it and creating an even larger cavity.
In this way, successive supernovae can overlap creating a network
of tunnels. Expanding at high speed within these tunnels,
the hot gas can move above the galactic plane where it is either halted 
by insufficient speed to escape the galactic potential, or by an encounter
with a large mass of cold, high density gas, or by efficient mixing
with cooler gas which increases its density thereby accelerating its
radiative energy losses.

In light of this complex environment in which star formation occurs,
it is even more surprising that the Schmidt law is so successful.
It is in the context of this complexity, that we undertake a study of the
star formation rate in a multiphase ISM. We restrict ourselves
to a local study of the ISM, namely that of a $\sim$ 1 kpc$^3$
region. The earliest local study which included supernovae feedback was
done by Rosen \& Bregman (1995) in two dimensions. They considered
a segment of a galactic disk, taking into account a fixed
external gravitational potential, but neglecting rotational effects,
self-gravity, and magnetic fields.  
In a three-dimensional model 
which included the effects of an external gravitational potential, rotation, 
and magnetic fields, Korpi et al. (1999a,b) studied a 
supernova driven galactic dynamo. Meanwhile, to investigate the disk
halo interaction, Avillez (2000) followed the evolution of a segment of a
galactic disk with an adaptive mesh refinement code.  Unlike these studies,
ours follows self-consistently and in three dimensions both the gas and the
stars, treating the latter as a system
of collisionless particles subject to gravity. Rosen \& Bregman (1995)
followed the stellar component but treated the stars with the same fluid
equations used for the gas thereby making their flow more viscous than that expected for
a collisionless system of particles.  Without star
particles tagged with their ages, Rosen \& Bregman (1995)
decided upon a supernovae rate for their simulation, then proceeded to
set off supernovae with a probability of occurrence correlated to the stellar density.
Avillez (2000) approached the issue by constructing an algorithm to
distinguish between isolated and clustered supernovae.
For isolated supernovae events, Avillez (2000) randomly determined the positions
of supernovae in the disk plane with
rates based on observed ones. To mimic clustered 
supernovae, a percentage of the supernovae sites were chosen to coincide with
locations where there was a previous supernova. In the Korpi et al. (1999a,b)
implementation there was a density criteria to determine the locations of
isolated supernovae. In both Avillez (2000) and Korpi et
al. (1999a,b), supernovae occuring above the disk plane were placed 
in random locations with an exponential distribution characterized by a scale
height also adopted from observations.
Given that we are interested in the impact of supernovae feedback on 
star formation, we cannot rely on these methods of modeling the supernovae
locations.  Instead we require that the locations, ages, and masses of the
star particles self-consistently determine the supernovae events.
A simple calculation shows that a star with a velocity of 10 km/s will travel
$\sim$ 100 pc (e.g. the average size of a molecular cloud) in 10 Myr. The latter 
corresponds to a typical time delay between the
birth and death of a star with M $\sim$ 80 M$_\odot$. In a follow-up paper
we explore how our results change when we neglect this time delay 
and instead allow the stars to explode as supernovae
immediately after their birth (Slyz, Devriendt, Bryan, \& Silk, 
{\em in preparation}). 
Obviously a local model such as ours
is of limited relevance for quantitative comparisons to the ISM in galaxies.
As later detailed in section~\ref{discussion}, the limitations of our idealized
boundary conditions and the absence in our models of an external gravitational
potential as well as of a shear flow arising from rotation 
means that there are many fundamental questions that we cannot 
address.  Nevertheless we believe that for the purposes of studying 
the non-linear interplay between star formation and stellar feedback, 
our simple model yields important insights. 

The question we address is what physical processes regulate the rate at which
gas turns into stars in a multiphase ISM. In section~\ref{method}
we describe the numerical method we use as well as the
ingredients of our simulation. To model the large dynamic range in 
densities and temperatures of a gaseous medium compressed by self-gravity and 
by shocks maintained by supernovae and stellar winds, a robust high-resolution 
hydrodynamical scheme proves essential.
To get a qualitative idea about the phenomena involved, 
section~\ref{generalpics} presents the general morphological, 
thermodynamical, and dynamical features of our simulations. 
A more quantitative
analysis of the gas structure and dynamics is presented
in section~\ref{detailedpics} where we explore changes in the gas probability
density function and energy spectra with the addition of more and more
physics thought to be relevant for star formation.
Section~\ref{comparesilk} compares the star formation rates we measure
in our simulations to simple analytic prescriptions
and section~\ref{discussion} discusses the limitations of our
simulations.
Finally our main conclusions are summarized in section~\ref{conclusion}.

\section{Method and Ingredients of the Simulations} 
\label{method}

Traditionally the problem with numerical simulations trying to model
star formation and feedback processes is that the radiative losses of
the hot component generated by supernovae are enourmous, even though
in the absence of any interaction of the hot gas with the cold gas the
cooling time of the hot gas is on the order of 100 Myr. In many cases
the culprit is numerical diffusion which mixes cold gas into 
the hot gas more than it physically should. As a result, since the 
density of the cold gas is high, mixing even a small fraction of it 
with the low
density hot gas increases the density of the hot component sufficiently
for it to cool more efficiently than it should. For this reason,
high resolution grid codes are better suited for studies of the 
multiphase interstellar medium than more diffusive particle based methods
which require carefully constructed algorithms to circumvent artificial 
cooling (e.g. Marri \& White 2003).

With this in mind, 
we model the evolution of gas and stars in a three--dimensional periodic box
which is 1.28 kpc on a side with a grid-based scheme for the gas and
a particle-mesh method for the stars. More specifically we have incorporated the BGK hydrocode
(Prendergast \& Xu 1993, Slyz \& Prendergast 1999) 
into Bryan's {{ENZO}} code (Bryan \& Norman 1997, 1999) which uses a
Lagrangean particle-mesh (PM) algorithm to follow the
collisionless stars moving in the gravitational potential 
the gas and the stars themselves generate. Based on gas-kinetic theory, BGK
computes time-dependent hydrodynamical 
fluxes from velocity moments of a distribution function which is a 
local solution to a model of the collisional Boltzmann equation, namely the
BGK equation (Bhatnagar et al. 1954).  The hydrodynamics code has been extensively tested 
on discontinuous non-equilibrium flows (see Xu 1998 for a review) and performs
well both at flow discontinuities and strongly rarefied regions, a criterion
which is mandatory for ISM simulations.

\begin{figure}
\centerline{\psfig{file=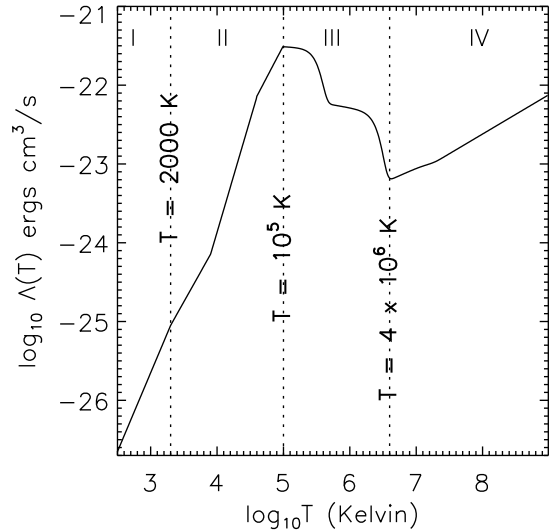,width=.9\hsize}}
 \caption{Cooling curve with vertical dotted lines overplotted to delineate
several different temperature regimes which we consider in section~\ref{generalpics}. }
\label{cool_function}
\end{figure}

Initially the gas has constant density ($\rho_{\mathrm {gas}} =$ 1 atom/cm$^3$) 
and temperature (T$_{\mathrm {gas}}=$ 10$^5$ Kelvin) and
similar to the initialization in MacLow et al. (1998),
its velocity field is drawn from a gaussian random field characterized by 
a power spectrum scaling like k$^{-4}$. We truncate this velocity 
power spectrum so that the field only has power on large scales, i.e. in modes 
up to k = 4.  The initial v$_{\mathrm{rms}}$ is
$\sim$ 50 km/s.
Contrary to MacLow et al. 1998, 
we do not add velocity perturbations at each time step to
drive the `turbulence`. We only impose the velocity
perturbations once at the beginning of the simulation. We assume radiative
cooling of an optically thin gas which is
in collisional ionization equilibrium. More specifically, 
our cooling function, displayed in figure~\ref{cool_function},
is an extension of the cooling curve of Sarazin \& White (1987)
down to temperatures of ${\mathrm {T}}_{\mathrm {min}} = 310$ K to 
account for ${\mathrm {H}}_{2}$ cooling using the rates given in 
Rosen \& Bregman (1995). The extension to lower temperatures assumes a solar 
metallicity, a completely ionized gas at 8000 K and an
ionization fraction that gradually drops to 10$^{-3}$ below 8000 K.
Fitting a piecewise power law to our cooling curve 
gives:
\begin{displaymath}
\Lambda(T) = \left\{ \begin{array}{ll} 0 & \mbox{if $T < 310 K$,} \\
(2.2380 \times 10^{-32}) T^{2} & \mbox{if $310 K \leq T < 2000 K$}\\
(1.0012 \times 10^{-30}) T^{1.5} & \mbox{if $2000 K \leq T < 8000 K$}\\
(4.6240 \times 10^{-36}) T^{2.867} & \mbox{if $8000 K \leq T < 39811 K$}\\
(3.1620 \times 10^{-30}) T^{1.6} & \mbox{if $39811 K \leq T < 10^5 K$}\\
(3.1620 \times 10^{-21}) T^{-0.2} & \mbox{if $10^5 K \leq T < 2.884 \times 10^5 K$}\\
(6.3100 \times 10^{-6}) T^{-3} & \mbox{if $2.884 \times 10^5 K \leq T  < 4.732 \times 10^5 K$}\\
(1.047 \times 10^{-21}) T^{-0.22} & \mbox{if $4.732 \times 10^5 K \leq T < 2.113 \times 10^6 K$}\\
(3.981 \times 10^{-4}) T^{-3} & \mbox{if $2.113 \times 10^6 K \leq T < 3.981 \times 10^6 K$}\\
(4.169 \times 10^{-26}) T^{0.33} & \mbox{if $3.981 \times 10^6 K \leq T < 1.995 \times 10^7 K$}\\
(2.399 \times 10^{-27}) T^{0.5} & \mbox{if $T \geq 1.995 \times 10^7 K$}
\end{array}
\right. 
\end{displaymath}
The lower temperature cutoff of the cooling 
function at 310 K is unphysical, although Rosen, Bregman, \& Norman
(1993) argue that truncating it there is a way to model the
contribution to the ISM pressure from sources such as magnetic fields
and cosmic rays, which do not decrease as the gas radiatively cools.

\subsection{Implementation of star formation and feedback}
Following Cen and Ostriker (1992), we assume that star formation 
is inevitable if a region is 
contracting ($\nabla \cdot v < 0$), cooling rapidly 
($\mathrm t_{\mathrm {cool}} < \mathrm t_{\mathrm {dyn}}$ and 
$T_{\mathrm {gas}} \leq T_{\mathrm {min}}$), 
and is overdense ($\rho > \rho_{\mathrm {crit}}$). Since we check the grid on a cell
by cell basis to see if these conditions are met, each timescale is computed
for each grid cell.  Here $t_{\mathrm {dyn}}$ is the
dynamical collapse timescale, i.e. $t_{\mathrm {dyn}} =  \sqrt{3.0 \pi/(32 G
\rho_{\mathrm{tot}})}$ where $\rho_{\mathrm{tot}}$ is the sum of the gas
density, $\rho$,
and the stellar density. $\;t_{\mathrm {cool}}$ is the cooling timescale,
i.e. $t_{\mathrm {cool}} = \mathrm{k T / n} \Lambda$, where $n$ is the 
gas particle number density. 
${\mathrm {T}}_{\mathrm {min}}$ is the minimum of our cooling curve, 310 K,  
and $\rho_{\mathrm {crit}}$ for the different simulations is specified in 
table~\ref{sims}.
If all our star forming criteria 
are met within a grid cell then we convert the following amount of gas, 
$\Delta m_{\mathrm {gas}} =  \epsilon \frac{\Delta t}{t_{\mathrm{dyn}}} 
 \rho_{\mathrm{gas}} \Delta x^3$ into a ``star particle'', where $\epsilon$ 
is a star formation efficiency
whose value is given in table~ref{sims}, and $\Delta t$ is the updating timestep. 
We only allow at maximum 90\% of the gas in a cell to be converted to stars in
one timestep.
In practice however, once supernovae inject hot gas into the
medium, the updating timestep is short as it is 
set by the hot, low density gas. As a result $\Delta t < t_{\mathrm{dyn}}$, and
this 90\% threshold is never reached.
We give the new star particle the same velocity as the gas out of which 
it formed and we follow the stars dynamically. 
The star particle is labeled with its mass, $m_{\star}$, its formation time, 
$t_{\mathrm{SF}}$, and the dynamical time, $t_{\mathrm{dyn}}$, of the gas out 
of which it formed.

For the purposes of the feedback however, rather than assume that
the ``star particle'' formed instantaneously at $t_{\mathrm{SF}}$, we 
spread the star formation over several dynamical times by
computing the amount of gas mass that would 
form stars after time $t_{\mathrm{SF}}$ to be:
\begin{equation}
    \Delta \mathrm{m}_{\mathrm{stars}}(t) = m_{\star} 
     \frac{(t - t_{\mathrm{SF}})}{\tau^{2}} {\mathrm{exp}}\frac{-(t - t_{\mathrm{SF}})}{\tau}
\label{expsfr}
\end{equation}
where $\tau = \mathrm{max}(t_{\mathrm{dyn}},10 \;{\mathrm {Myr}})$.
With this time-dependent star formation rate, stars form at an exponentially
decreasing rate after a dynamical time. If the dynamical timescale 
of the gas in a
star-forming cell is shorter than the typical lifespan of a massive star,
i.e. 10 Myr, then 10 Myr is used in place of $t_{\mathrm{dyn}}$ in
equation~\ref{expsfr} for the value of $\tau$.
Then, as a crude model of a stellar wind, we return 
25\% of $\Delta \mathrm{m}_{\mathrm{stars}}$
to the gas, and since this returned mass
has the velocity of the ``star particle'' we alter the momentum
of the gas appropriately.
Finally assuming only the occurence of Type II supernovae, 
we add 10$^{-5}$ of the rest-mass energy of 
$\Delta \mathrm{m}_{\mathrm{stars}}$
to the gas' thermal energy (Ostriker \& Cowie 1981). The
supernovae input is added locally into one cell. We explore the limitations
of our supernovae implementation in future work.
As we do not have the resolution to follow every individual
star and to therefore sample a realistic Initial Mass Function (IMF) 
for them, each star particle is more like a small star cluster
with a typical mass in the range $\sim 120 - 220 {\mathrm M_{\odot}}$.

Table~\ref{sims} presents the simulations we ran, listing the values
of the parameters for star formation and feedback.
Although we performed several simulations with a density threshold for star
formation, $\rho_{\mathrm {crit}}$, set to 1 atom/cm$^{3}$ (runs B5, C5 
and C6),
for the remainder of the paper we focus only on the runs with 
$\rho_{\mathrm {crit}}$ = 10 atom/cm$^{3}$.   This is because we found that
dropping the density threshold by one order of magnitude to 
1 atom/cm$^{3}$ did not change the SFR by a factor ten, but merely by
about 10\% at the peak of star formation.  As Table~\ref{sims} indicates,
we also experimented with the value of $\epsilon$ and found that taking
a value of $\epsilon$ = 0.01 (ten times smaller than our fiducial value)
left the conclusions presented in this paper unchanged, i.e. the medium
became porous and the SFR peaked at roughly the same value although
with a slight time delay compared to the run with $\epsilon$ = 0.1 .

\begin{table*}
\caption{Summary of the performed runs. All of the runs have radiative cooling.
The first three columns indicate 
whether self-gravity, star formation and/or feedback are activated. 
$\rho_{\mathrm {crit}}$ is the density threshold for star formation,
$\epsilon$ is the star formation efficiency and the final column 
indicates the grid resolution. Each simulation cube is 1.28 kpc per side.
}
\label{sims}
 \begin{tabular}{l||cc|cc|cc|cc} 
    & self--gravity & stars & feedback  &  $\rho_{\mathrm {crit}}$
 (at/${\mathrm {cm}}^3$)  & $\epsilon$  & grid resolution (pc) \\ \hline \hline
 A    &      --      &     --      &   --      &  --    & --   &  5   \\
 B1   &      --      &    yes      &   --      &  10.   & 0.1  &  10  \\
 B2   &     yes      &    yes      &   --      &  10.   & 0.1  &  10  \\
 B3   &      --      &    yes      &  yes      &  10.   & 0.1  &  10  \\
 B4   &     yes      &    yes      &  yes      &  10.   & 0.1  &  10  \\
 B5   &     yes      &    yes      &  yes      &   1.   & 0.1  &  10  \\
 B6   &     yes      &    yes      &  yes      &  10.   & 0.01 &  10  \\
 C1   &      --      &    yes      &   --      &  10.   & 0.1  &  20  \\
 C2   &     yes      &    yes      &   --      &  10.   & 0.1  &  20  \\
 C3   &      --      &    yes      &  yes      &  10.   & 0.1  &  20  \\
 C4   &     yes      &    yes      &  yes      &  10.   & 0.1  &  20  \\
 C5   &     yes      &    yes      &  yes      &   1.   & 0.1  &  20  \\
 C6   &     yes      &    yes      &  yes      &   1.   & 0.01 &  20  \\
\\
\\
 \end{tabular}
\end{table*}

\section{General Features of the Multiphase Medium}
\label{generalpics}
\begin{figure*}
\centerline{\psfig{file=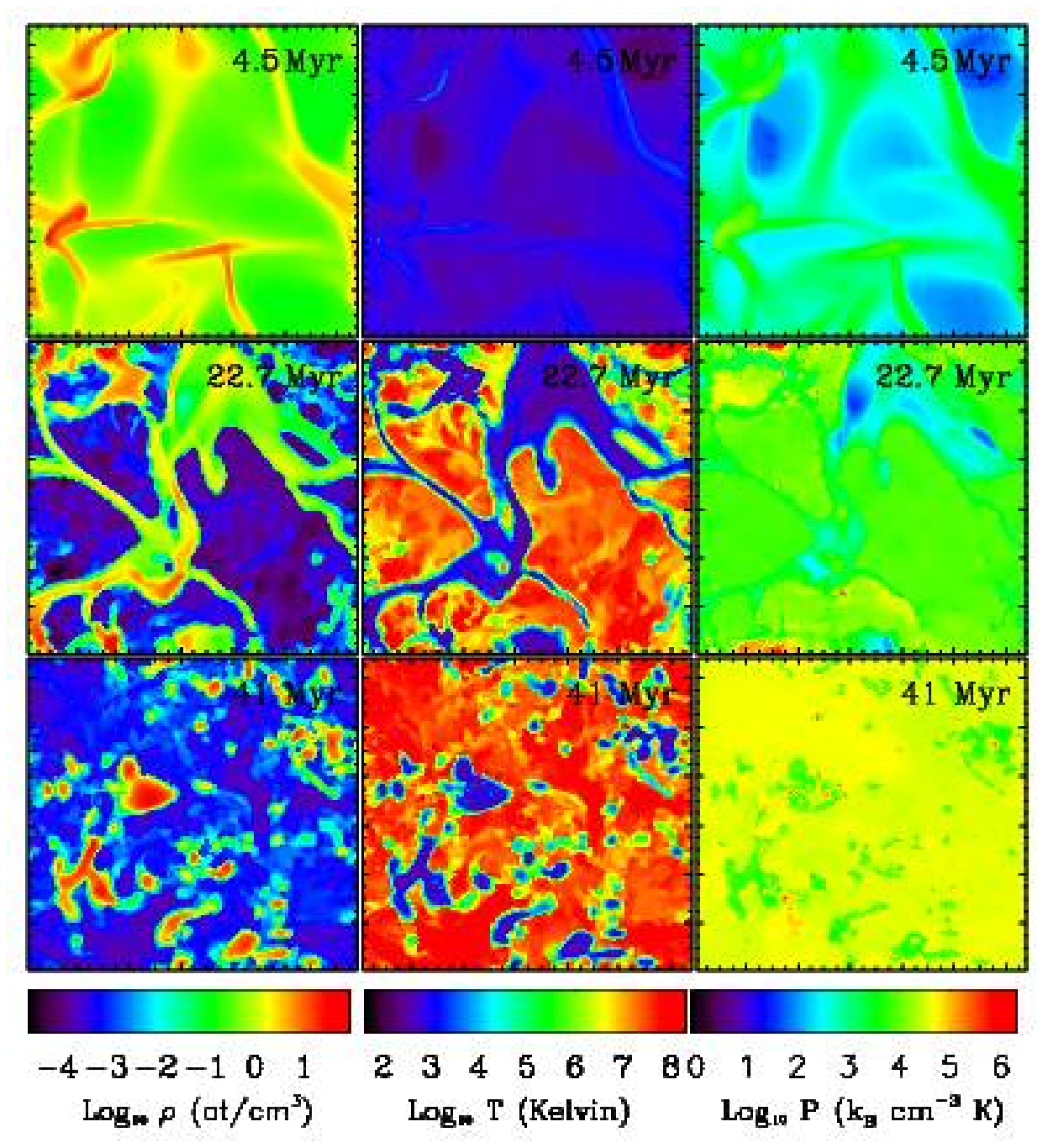,width=\hsize,angle=270}}
 \caption{Time evolution of the logarithm of the gas density (first column),
   temperature (second column) and pressure (third column) in a 12.8 pc
$\times$ 1.28 kpc $\times$ 1.28 kpc slice for run B4.}
\label{dTp_128}
\end{figure*}

We begin by showing the time evolution of one of our simulations, namely B4, 
which includes all the physical processes we considered, namely 
``turbulent'' initial conditions (as defined in section~\ref{method}), radiative cooling, self-gravity, 
star formation and feedback.  In figure~\ref{dTp_128}  
we show the gas density, temperature and pressure
in a 12.8 pc $\times$ 1.28 kpc $\times$ 1.28 kpc slice 
of this run. Due to the compression caused by 
turbulence and self-gravity, the gas in certain regions, 
satisfies our criteria for star formation. Following their formation,
this first generation of stars soon explodes as 
supernovae, releasing hot gas into the interstellar medium.
The morphologies of the hot bubbles are extremely non-spherical
due to the fact that the supernovae are releasing their thermal energy
into a spatially inhomogeneous and non-stationary medium.
Because this hot, low density gas has a long cooling time and because 
the star formation rate is sufficiently high, subsequent generations 
of supernovae bubbles overlap, filling more and more of the volume.  
Ultimately the density and temperature span more than six orders of 
magnitude in such a simulation and are anti-correlated: 
high density regions are cold, and low density regions are hot.
As the third column in figure~\ref{dTp_128} shows, this anti-correlation results 
in near pressure equilibrium between these two phases of the gas.
Nevertheless the dense gas is about one order of magnitude lower in pressure
than the low density gas indicating that a thermal instability is active.
Other regions which are out of pressure equilibrium by 1 -- 2 orders of
magnitude are  those which have just experienced 
thermal energy input from supernovae. Self-gravitating gas would also appear 
out of pressure equilibrium, something we see in
later stages of the simulation.

\begin{figure}
\centerline{\psfig{file=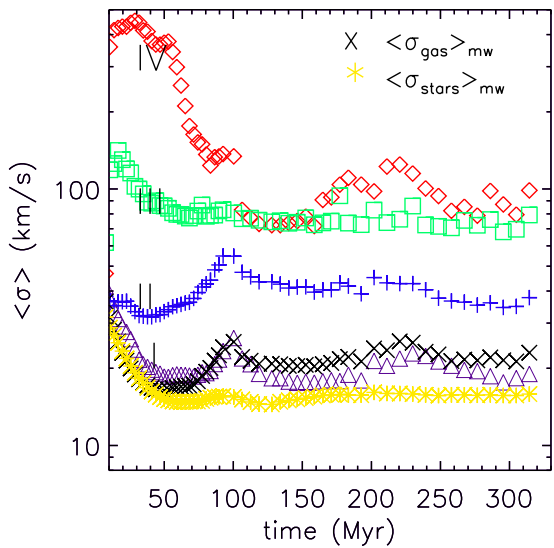,width=.9\hsize}}
 \caption{Time evolution of the logarithm of the velocity 
dispersion in run B4 for the gas in different temperature
regimes: (I) T $<$ 2000 K (triangles), (II) 2000 K $<$ T $<$ $10^5$ K 
(plus signs), (III)$10^5$ K $<$ T $<$ $4 \times 10^6$ K (squares), 
(IV) $4 \times 10^6$ K $<$ T (diamonds). The crosses mark the
average mass--weighted velocity dispersion of the gas and the asterices that
of the stars.}
\label{sigma}
\end{figure}

The dynamical state of the stars and of the gas in different temperature 
regimes in the simulation is summarized by a plot of the average velocity 
dispersions (fig.~\ref{sigma}). Guided by some of the features in the cooling
curve (see figure~\ref{cool_function}), we divide the temperature into
the following four categories: (I) T $<$ 2000 K, 
(II) 2000 K $<$ T $<$ $10^5$ K,
(III)$10^5$ K $<$ T $<$ $4 \times 10^6$ K, (IV) $4 \times 10^6$ K $<$ T.
We compute the average velocity dispersion of the gas in each of these
4 regimes, and in addition, we calculate the mass-weighted velocity
dispersion of the gas, as well as the mass-weighted velocity dispersion 
of the stars.
As the stars are assigned the velocity of their progenitor gas at formation,
their velocity dispersion closely follows the velocity dispersion of the cold
gas. Furthermore, we find that with the exception of the hottest phase (IV),
the velocity dispersion of the other phases approximately settles to
the following values: (I) 15 km/s, (II) 30 km/s and (III) 75 km/s.
What is very striking in the plot of
the velocity dispersions, is the high velocities ($\sim$ 500 km/s) 
attained by the hot, low density component of the gas. 
The densest structures which provide the raw material for star formation,
collide and break apart, but are also subject to stripping via 
hydrodynamical and thermal instabilities when 
this hot, low density material flows rapidly past them.
The picture of a ``violent interstellar medium'' (McCray \& Snow 1979) emerges.

\begin{figure*}
\centerline{\psfig{file=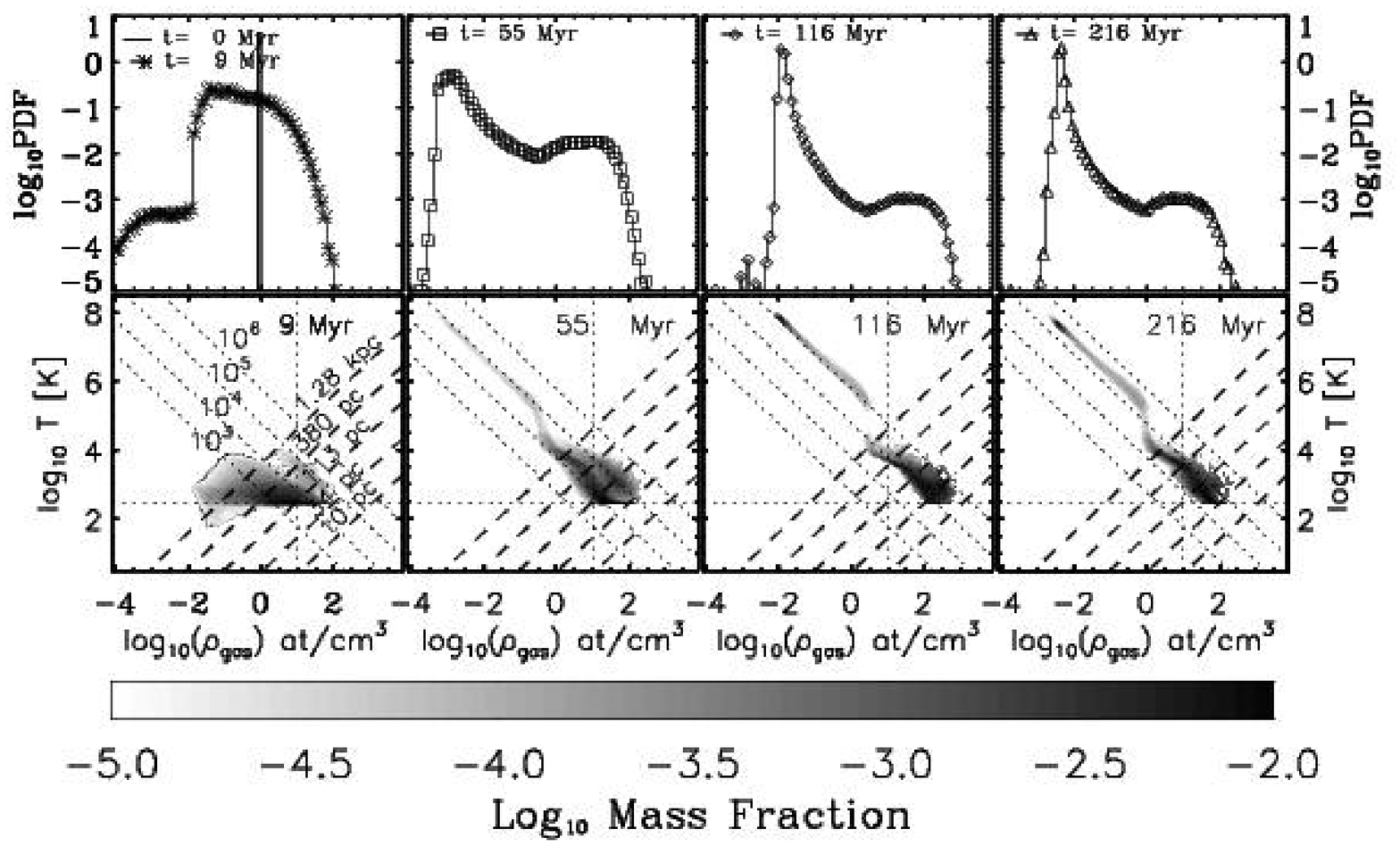,width=.6\hsize,angle=270}}
 \caption{Time evolution of the density PDF (top row) and phase diagrams 
(bottom row) for run B4 (128$^3$ run with star formation, feedback and 
self-gravity). 
In the phase diagrams, the dotted vertical (horizontal) line marks the critical density, $\rho_{\mathrm {crit}}$,  (temperature, $T_{\mathrm {crit}}$)
for star formation. Dotted diagonal lines mark lines
of constant pressure, and are labeled for the ${\mathrm t} = 0$ Myr frame:
${\mathrm P} = 10^{6}$, $10^{5}$, $10^{4}$,
and $10^{3}\;{\mathrm k}_{\mathrm B} \;{\mathrm {cm}}^{-3}\; {\mathrm K}$. 
Dashed diagonal
lines (labeled for the $t=0$ Myr frame) mark the Jeans length:
$\lambda_J = $10 pc, 34 pc, 113 pc, 380 pc and 1.28 kpc. }
\label{thermalevolution_4}
\end{figure*}

Regarding the evolution of the thermal state of the gas, this is
well portrayed in phase diagrams of the gas (bottom row of 
figure~\ref{thermalevolution_4}) which show the distribution of the mass
fraction of the gas as a function of its temperature and density.
Given our initial conditions of uniform density and temperature, if we were
to plot a phase diagram of the gas at time $t = 0$ Myr, all the 
gas would occupy a single point.
Because the initial temperature ($10^5$ K) 
of the gas coincides with the peak of the cooling curve, by 
9 Myr (first panel of bottom row of figure~\ref{thermalevolution_4}) the 
majority of the gas quickly radiatively cools to an approximately isothermal
state at a temperature corresponding to the minimum of the cooling curve,
i.e. 310 K. As we instantaneously imprint a spectrum of velocity perturbations
at the beginning of the simulation, the gas acquires a range of density values
and therefore has a spread in densities by this time. 
Thereafter, with the injection of hot gas into the medium, a tail of low density, hot 
gas appears.  
However as gas with temperatures $10^5$ K $<$ T $<$ $4 \times 10^6$ K (phase
III) is thermally unstable, it gradually vanishes from the medium, 
dividing the gas into two parts in the phase diagram. 
The majority of the coldest (T $\sim$ 300 K) gas 
differs by  approximately a one order
of magnitude pressure jump from gas with T $\geq$ 5 $\times$ $10^5$ K. 
Finally the pressure of both the hot and cold gas changes with time.
It rises as more and more hot gas fills the simulation volume,
a situation that would probably be different if hot gas were allowed
to escape the box.

\begin{figure*}
\centerline{\psfig{file=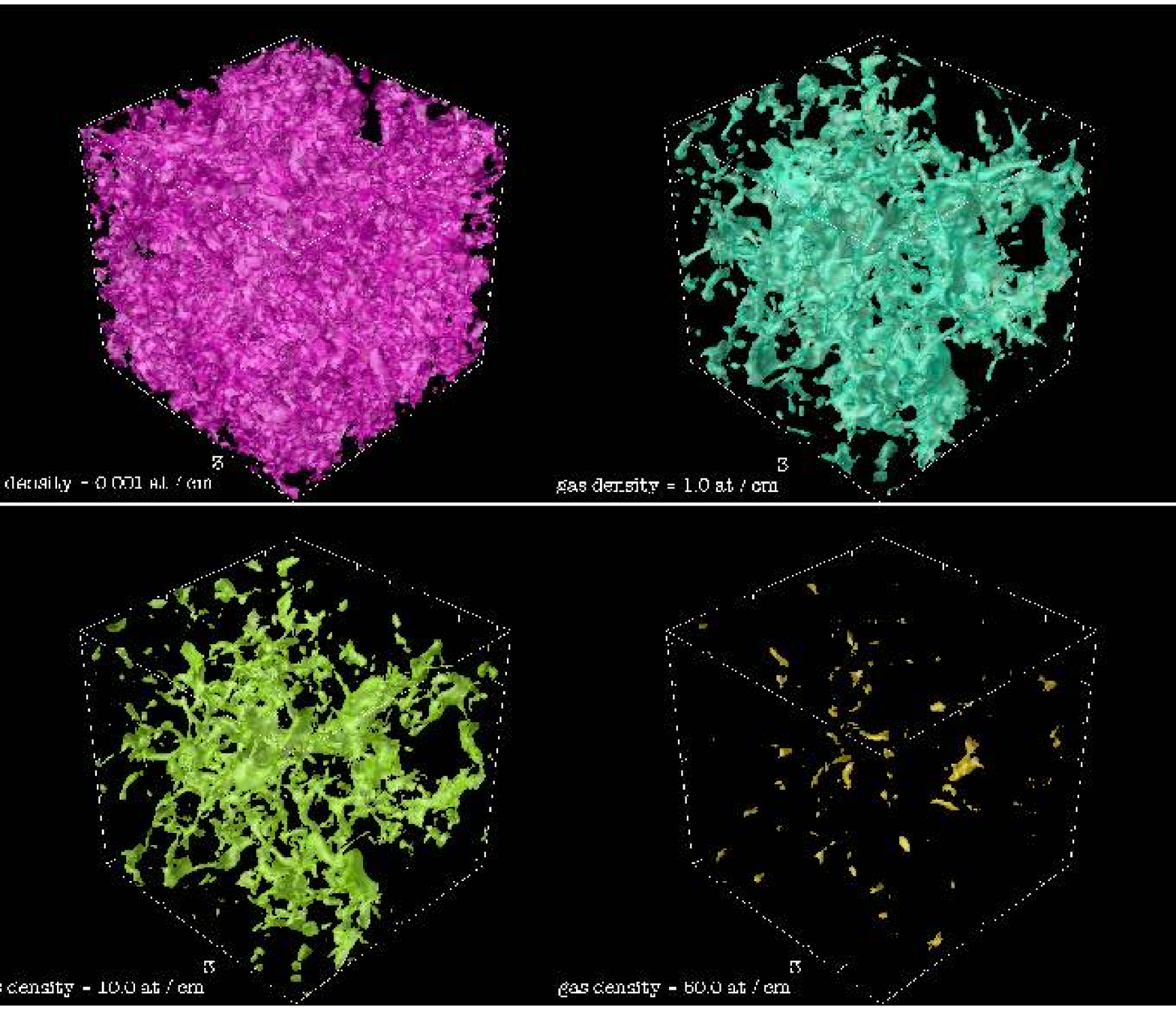,width=.7\hsize,angle=180}}
 \caption{Isodensity surfaces of the gas for run B4 and $\rho = 10^{-3}, 1, 
10,$ and 50 atoms/${\mathrm {cm}^3}$.}
\label{gas3d}
\end{figure*}
\begin{figure}
\centerline{\psfig{file=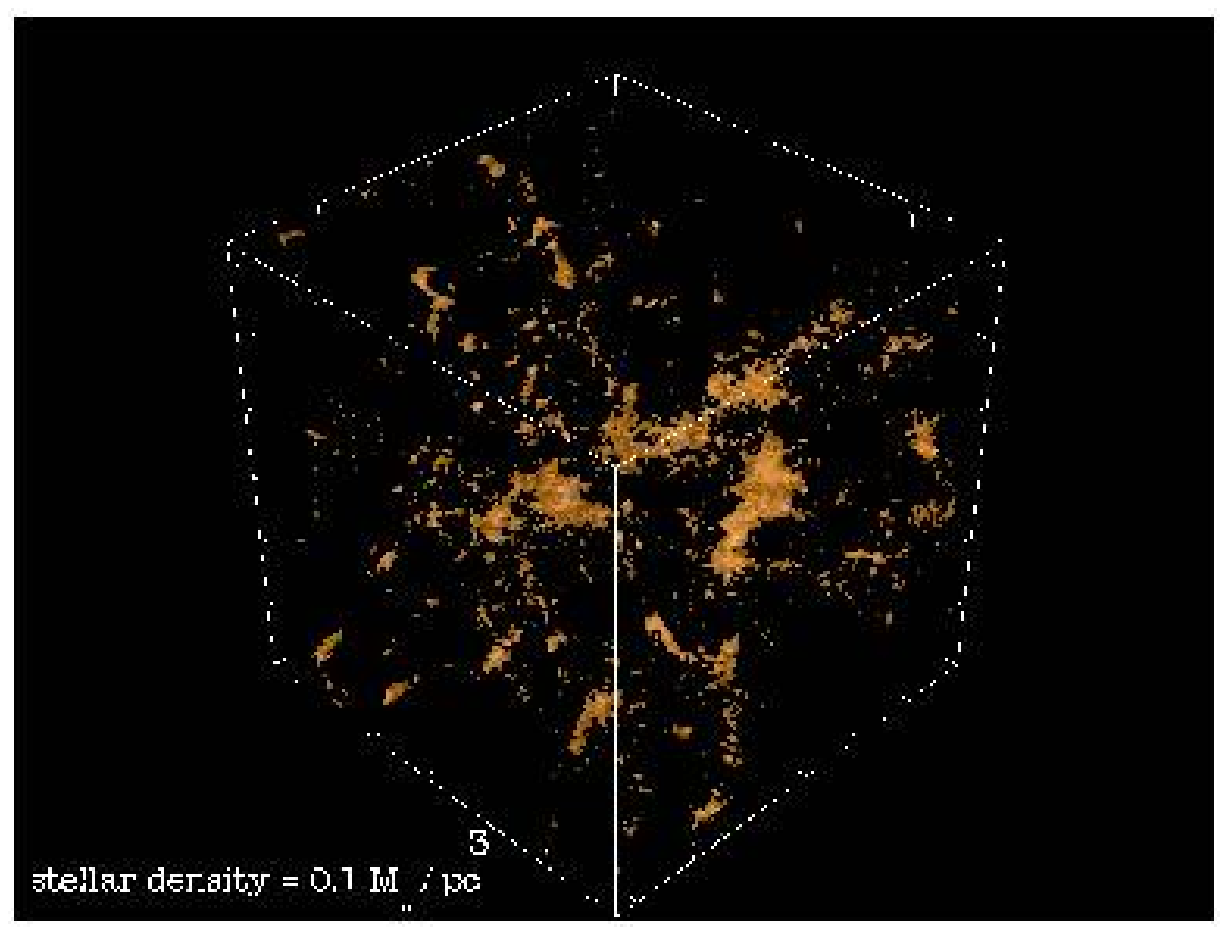,width=.75\hsize,angle=270}}
 \caption{Isodensity surface of the stellar density for run B4 and 
$\rho_{\star} = 0.1$ M${_{\odot}/\mathrm {pc}^3}$.}
\label{star3d}
\end{figure}

Although complex, pictures of the gas density and temperature distribution 
in a two-dimensional slice through the simulation volume, do not
capture the intricacy of the three-dimensional structure. In an attempt
to display this structure, in figure~\ref{gas3d} we plot isodensity
surfaces of the gas for $\rho = 10^{-3}, 1, 10,$ and 50
atoms/${\mathrm {cm}^3}$ at 50 Myrs. It is clear from these figures that 
the hot, low density
component fills most of the volume, while the densest regions fill the 
smallest fraction of the space, and are scattered throughout the box.
A three-dimensional rendering  of the stellar density at the same time
instant (fig.~\ref{star3d}), reveals traces of the imprint of the high density gas 
distribution and encouragingly bears some qualitative resemblance to the  
distribution of H$\alpha$ emission in disk galaxies 
(e.g. NGC 4631, Wang et al. 2001).

\section{Quantifying the Structure and Energetics of the Multiphase Medium}
\label{detailedpics}

\begin{figure*}
\centerline{\psfig{file=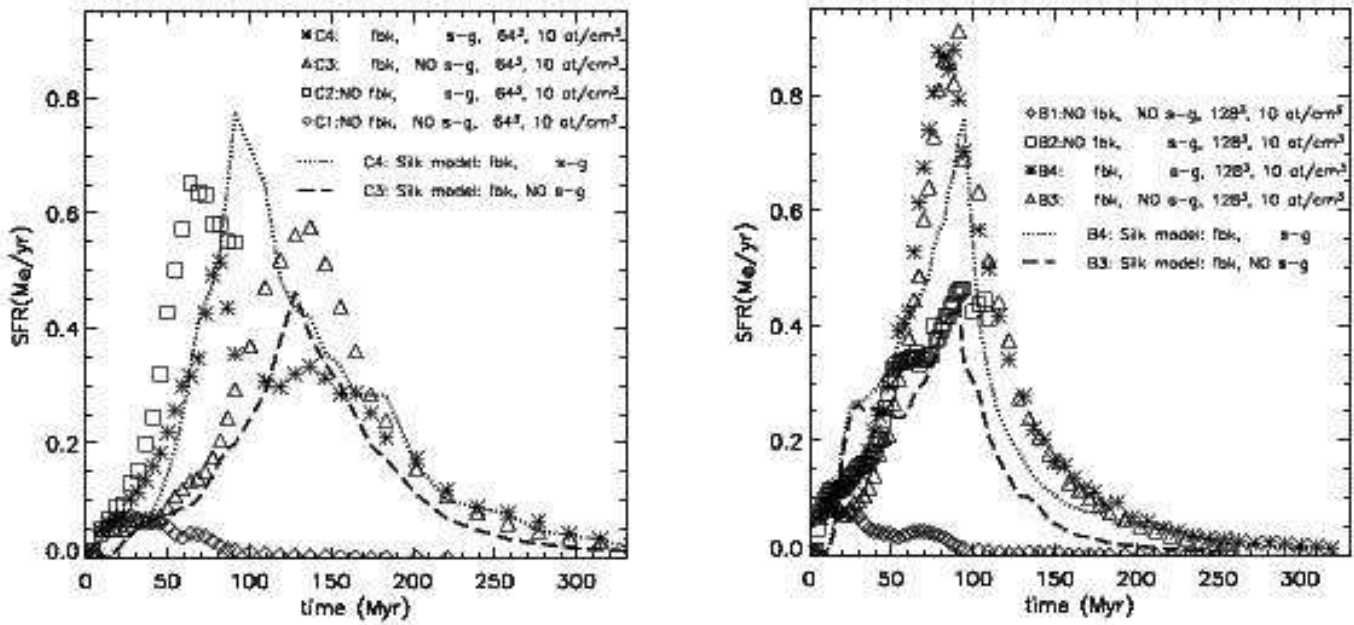,width=.5\hsize,angle=270}}
 \caption{Time evolution of the star formation rate
for a series of runs (see table~\ref{sims}) differing in their physics.
The left panel displays the results from runs C1 (diamonds), C2 (squares),
C3 (triangles) and C4 (asterices). The right panel displays the results
from runs B1 (diamonds), B2 (squares), B3 (triangles) and B4 (asterices).
Symbols are the measured SFRs and the dotted and dashed lines are
analytic models from Silk (2001).}
\label{sfrs}
\end{figure*}

In an effort to assess what determines star formation rates, 
we systematically examine how different physical processes change 
the structure and the energetics of the interstellar medium.
The sequence of runs listed in table~\ref{sims} are designed to
isolate the effects of successively more complicated physical processes.
A plot comparing the star formation rates for this sequence of runs 
(figure~\ref{sfrs})
invites us to study what keeps star formation at a minimum and alternatively
what is necessary to drive it to high values.
Resolution effects immediately manifest themselves in figure~\ref{sfrs}.
The $64^3$ and $128^3$ runs start from the same initial conditions. Preceeding
star formation, feedback is non-existent, but self-gravity plays a larger
role in the $64^3$ run where a grid cell of equivalent density to that in the
$128^3$ grid is 8 times more massive. Therefore in the $64^3$ run with only
self-gravity (run C2), the SFR rises more rapidly at
earlier times than for the comparable run performed on the $128^3$ grid (run
B2). Once feedback occurs, a mechanism
supplementary to turbulence exists for creating high density contrasts which
are stronger in the higher resolution runs. This causes higher peaks of SFR
in the $128^3$ runs with feedback (runs B3 and B4) as compared to
the equivalent $64^3$ runs (C3 and C4). On the other hand, feedback also
creates an extra source of pressure to fight self-gravity which explains
why the C2 run leads to higher SFRs at earlier times than the C3 and C4 runs.
What remains unclear without performing a simulation at
still higher resolution is whether the indistinguishability between the
$128^3$ runs
with feedback regardless of whether or not there is self-gravity (runs B3 and
B4) are a manifestation of convergence or coincidence. However, we believe
convergence is the more probable explanation as increasing the resolution tends to increase
the dominance of feedback processes over self-gravity.  More specifically,
in the case of the $64^3$ runs a rise in the SFR is driven more rapidly when
self-gravity is included. In contrast, star formation increases at similar rates
regardless of whether self-gravity is included in the $128^3$
runs.  Therefore we do not see any reason why this trend should be inverted by
further increasing the resolution.

Before proceeding, we calculate roughly the supernovae rate
corresponding to the measured star formation rates in our
simulations. In our $1.28^3$ ${\mathrm {kpc}^3}$ box, typical star
formation rates are SFR $\sim 0.1 - 0.8 \,{\mathrm M}_\odot/{\mathrm {yr}}$. Scaling
these values to a Milky Way type galaxy, where $M_{\mathrm {MW}}$ is the
mass of gas in the Milky Way, and $M_{\mathrm {box}}$ is the mass of
gas in our simulation cube, 
\begin{equation}
{\mathrm {SFR}} \, (M_{\mathrm {MW}}/M_{\mathrm {box}}) \approx 100 - 800
\,{\mathrm M}_\odot/{\mathrm yr}.
\end{equation} 
For a Salpeter IMF there is approximately 1 SN/200 ${\mathrm M}_\odot$,
implying that the typical supernovae rates in our simulation volume
are $\sim 0.5 - 4$ SN/yr. Furthermore, with this scaling to higher mass 
the projected gas surface density increases
by about 4 orders of magnitude bringing both the SFRs and 
surface densities to values representative of the starburst regime in the Kennicutt relation
(Kennicutt 1998).

\begin{figure*}
\centerline{\psfig{file=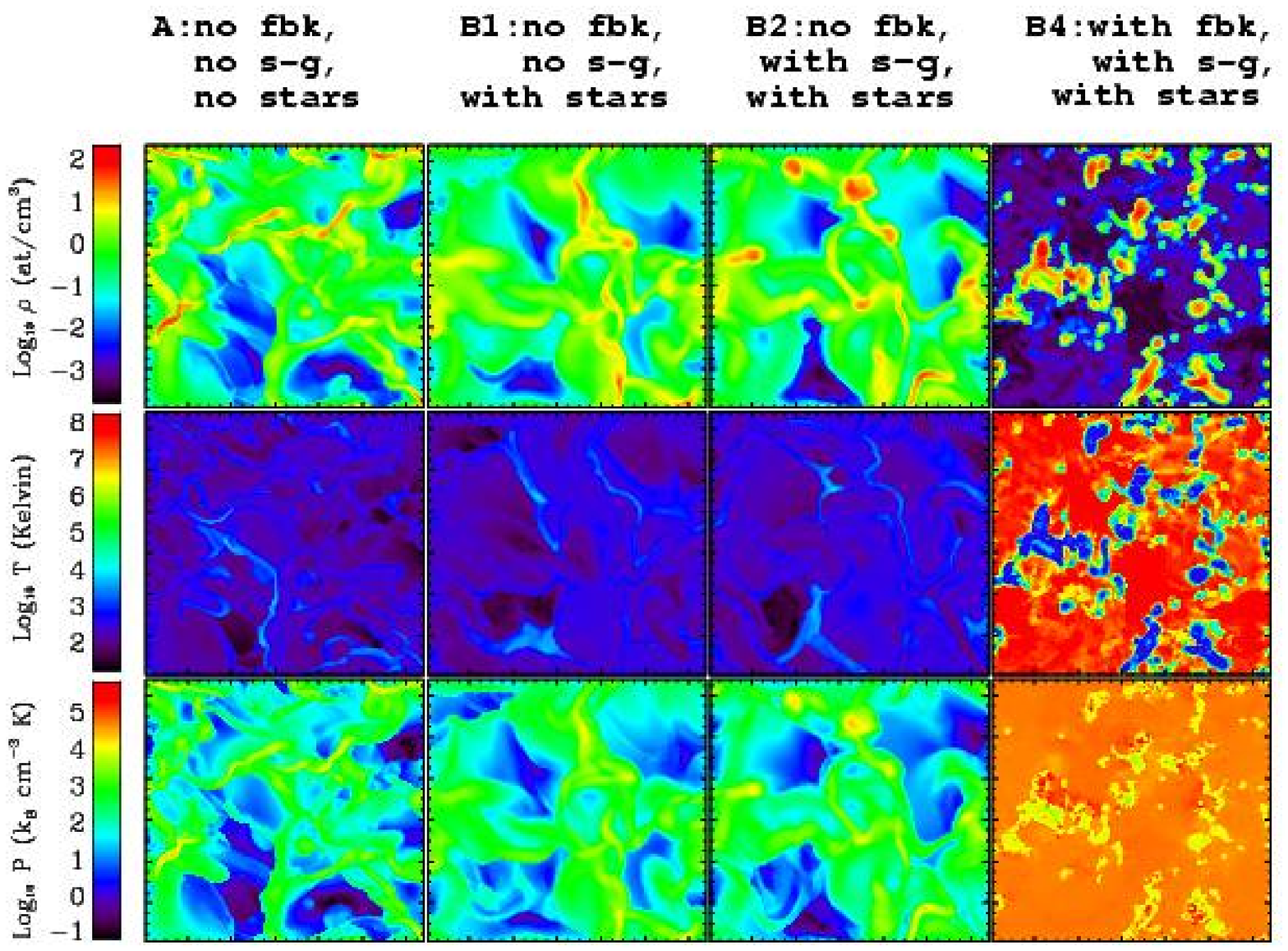,width=.85\hsize,angle=270}}
 \caption{The gas density, temperature and pressure at time 
${\mathrm t = 45}$ Myrs in a 
12.8 pc $\times$ 1.28 kpc $\times$ 1.28 kpc slice for runs A, B1, B2 and B4
(see table~\ref{sims} for the specifications of each of these runs). }
\label{compare_rhoTp}
\end{figure*}
A visual examination of a 2D snapshot of the gas density, 
temperature and pressure taken at the same time (${\mathrm t = 45}$ Myrs)
for runs including different physics is useful for comparing
some of the consequences of the different processes. 
Figure~\ref{compare_rhoTp} clearly shows how self-gravity, which
is a radially directed force towards regions of locally high density,
causes high density regions to take on a more spherical
appearance. Furthermore, all the runs without feedback have
gas with pressure spanning over $\sim$ 6 orders of magnitude, and
a small range in temperatures compared to the run with feedback.
The low density gas regions in the runs without feedback, are cold (T $\sim$ 300 K)
and are created by adiabatic cooling during extreme expansion
in certain regions of the ``turbulent'' medium. Another feature
that appears in this sequence of simulations is that the dense structures
in the run with feedback are sharper due to destruction of intermediate
density material by thermal and hydrodynamical instabilities. 
\subsection{Probability Density Function of Mass Density}

A density probability distribution function (PDF) is a simple 
one-dimensional statistical 
measure of the structure of a medium. In practice for simulations performed
on a grid, PDFs are instantaneous 
histograms tallying the number of grid cells of a certain density in the 
simulation. Under the premise that stars form in high density regions, the
statistical properties of the density field, itself nonlinearly coupled
to the velocity field, might give clues to the process of star formation.
Efforts to uncover how the gas density organizes itself in media 
structured by different dynamical processes are ongoing. 
V\'{a}zquez-Semadeni (1994) presented a statistical argument to show that 
turbulent (random), supersonic, compressible flows
naturally generate hierarchical structure without necessitating an appeal
to things like fragmentation in a gravitationally unstable system (Hoyle 1953).
In the limit of very high Mach numbers these flows have a pressureless 
behaviour and if, in addition, self-gravity is negligible then the 
hydrodynamical equations are scale-invariant. Consequently, whatever the 
density in a given region, that region has the same probablity of producing 
a relative fluctation with respect to its normalizing density, as any other 
region in the flow. Assuming that in a random flow successive density steps are
independent, the central limit theorem dictates that the density distribution
should be log-normal. And indeed, V\'{a}zquez-Semadeni's (1994) 
two-dimensional, essentially isothermal
($\gamma$ = 1.0001) simulations of a weakly compressible (M $\sim$ 1), 
turbulent flow without self-gravity developed log-normal density PDFs both on
the large scale of the simulation and in subregions within the simulation.

Subsequently, numerical experiments of three-dimensional, isothermal, 
randomly forced, supersonic turbulence by Padoan, Nordlund \& Jones (1997) 
also found that the gas density follows a log-normal distribution,
\begin{equation}
{\mathrm {PDF}} = \, \frac{1}{\sigma \sqrt{2 \pi}} \, {\mathrm e}^{-({\mathrm {ln}}\rho - <{\mathrm {ln}}\rho>)^2/2\sigma^2}.
\end{equation}  
Furthermore they observed empirically that 
the dispersion, $\sigma$, of the log--normal scales with the
root--mean--squared Mach number, $M_{\mathrm {rms}}$, as follows:
\begin{equation}
\sigma^2 = \, {\mathrm {ln}}(1 + (\frac{M_{\mathrm {rms}}}{2})^2)
\end{equation}  
or, for the case of the linear dispersion
\begin{equation}
\sigma_{\mathrm{linear}} = \, \frac{M_{\mathrm {rms}}}{2}.
\end{equation} 
These dispersion relations reflect the fact that in a medium with
higher $M_{\mathrm {rms}}$, the gas achieves greater density contrasts.
Passot \& V\'{a}zquez-Semadeni (1998) found the same linear scaling
relation for the isothermal case. A formal proof for the lognormal
PDF in the case of isothermal, supersonic turbulence was provided
by Nordlund \& Padoan (1999) based on the formalism given in 
Pope \& Ching (1993).
\begin{figure*}
\centerline{\psfig{file=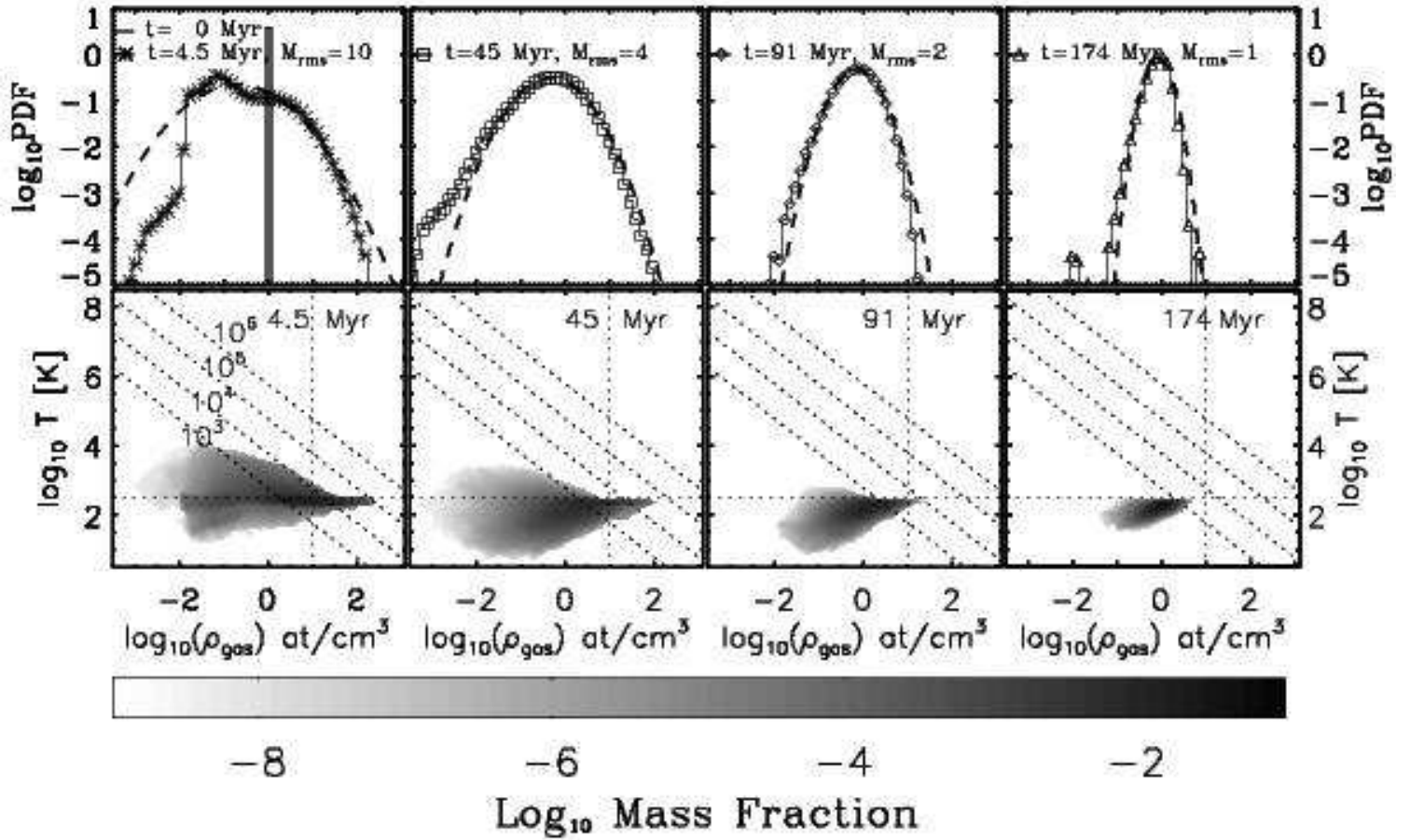,width=.7\hsize,angle=270}}
 \caption{Time evolution of the density PDF (top row) and phase diagrams 
(bottom row) for run A (128$^3$ run with no star formation, no feedback
and no self-gravity). The thick dashed line overplotted on the measured 
PDFs (symbols) is the log--normal PDF predicted by Padoan, Nordlund \& Jones (1997).
In the phase diagrams, the dotted vertical (horizontal) line marks the critical density, $\rho_{\mathrm {crit}}$,  (temperature, $T_{\mathrm {crit}}$)
for star formation. Dotted diagonal lines mark lines
of constant pressure, and are labeled for the ${\mathrm t} = 0$ Myr frame:
${\mathrm P} = 10^{6}$, $10^{5}$, $10^{4}$,
and $10^{3}\;{\mathrm k}_{\mathrm B} \;{\mathrm {cm}}^{-3}\; {\mathrm K}$.}
\label{pdf_nofbknosg}
\end{figure*}

Scalo et al. (1998) and Passot \& V\'{a}zquez-Semadeni (1998) 
extended this work on isothermal flows by considering the polytropic case.
Having conducted two-dimensional simulations including various combinations
of physical processes (e.g. self-gravity, magnetohydrodynamics, Burgers
turbulence),  Scalo et al. (1998) found PDFs that were more consistent
with power laws than with log-normal distributions. Seeking to understand
this result and its discrepancy with previous work on isothermal
flows which consistently found lognormal distributions, 
Scalo et al. (1998) performed one-dimensional simulations of forced, 
supersonic, polytropic turbulence
and uncovered a lognormal PDF for the cases where either the gas was isothermal
($\gamma$ = 1) or where the Mach number was small (M$\ll$ 1).
Otherwise, when $\gamma <$ 1, power laws developed for densities 
larger than the mean. Alternatively,  
Nordlund \& Padoan (1999) interpreted Scalo et al.'s results for
the PDFs occuring in the $\gamma \neq$ 1 case as skewed log-normals
and Passot \& V\'{a}zquez-Semadeni (1998) provided a mathematical framework
for understanding why these distributions arose.  

\begin{figure*}
\centerline{\psfig{file=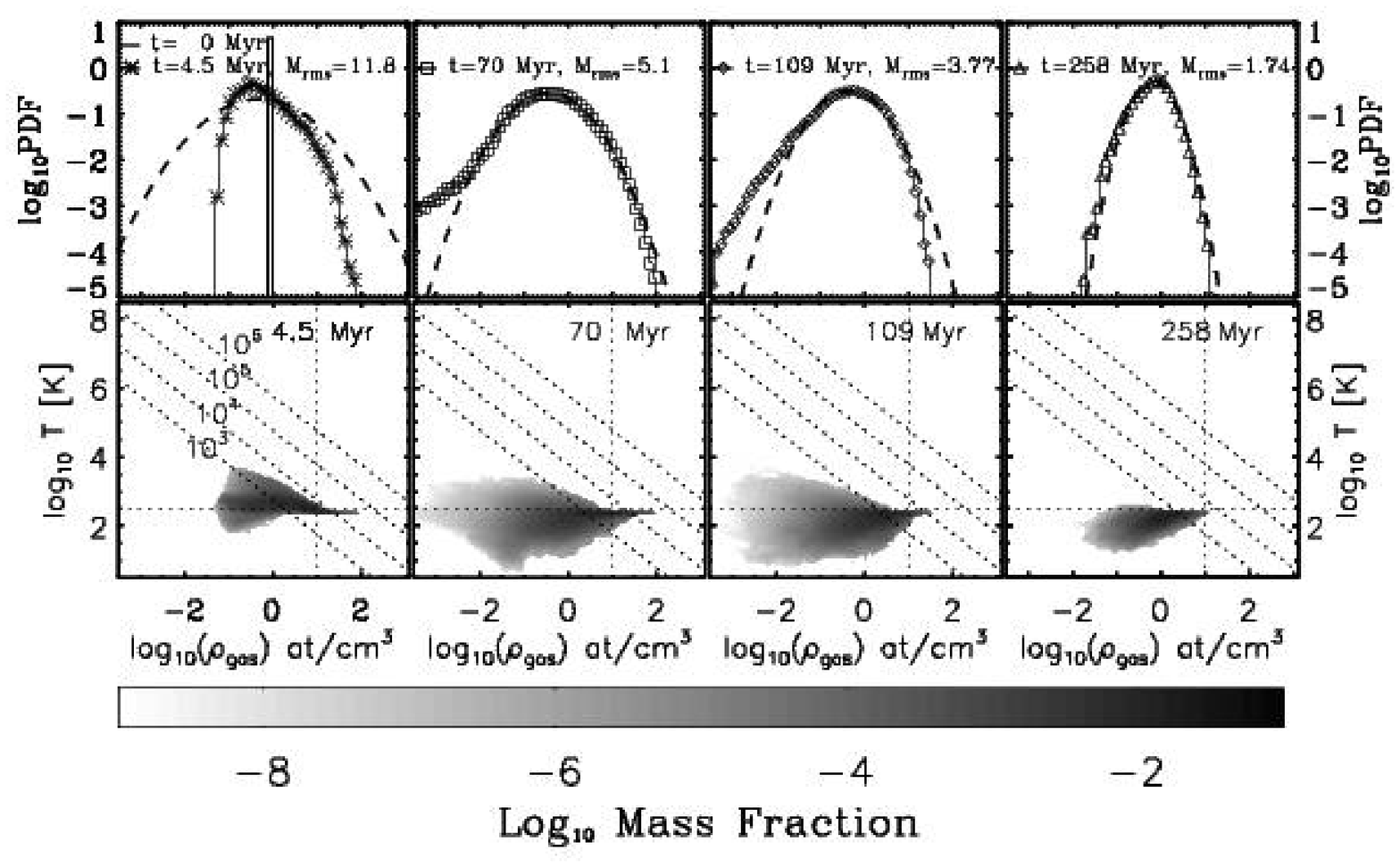,width=.7\hsize,angle=270}}
 \caption{Time evolution of the density PDF (top row) and phase diagrams 
(bottom row) for run B1 (128$^3$ run with star formation, no self-gravity, and
no feedback). The thick dashed line overplotted on the measured 
PDFs (symbols) is the log--normal PDF predicted by Padoan, Nordlund \& Jones (1997).
In the phase diagrams, the dotted vertical (horizontal) line marks the 
critical density, $\rho_{\mathrm {crit}}$,  (temperature, $T_{\mathrm {crit}}$)
for star formation. Dotted diagonal lines mark lines
of constant pressure, and are labeled for the ${\mathrm t} = 0$ Myr frame:
${\mathrm P} = 10^{6}$, $10^{5}$, $10^{4}$,
and $10^{3}\;{\mathrm k}_{\mathrm B} \;{\mathrm {cm}}^{-3}\; {\mathrm K}$}
\label{pdf_nofbknosgstars}
\end{figure*}

Our work extends these investigations on the PDF in the direction
of the cases where the ISM is constrained neither to be isothermal 
nor polytropic.
As a result our local temperature and pressure are not simple functions of the
density but arise from the evolution of the thermal energy.
Because we consider processes (e.g. radiative cooling, self--gravity, 
star formation) whose effectiveness depends on the density, the hydrodynamic
equations are no longer scale--invariant. Therefore the condition of 
randomness between subsequent density fluctuations is violated and 
one cannot expect a log--normal density PDF 
(e.g. V\'{a}zquez-Semadeni's (1994)). In our series of experiments of 
increasing complexity (see Table~\ref{sims}), the simplest
simulation we performed was of non--isothermal supersonic turbulence (run A).
Despite the inclusion of density--dependent cooling processes, we found
that the structure of the gas quickly evolved to a density PDF consistent
with a log--normal. This is not a surprising result since without a heat source
the majority of the gas quickly cools to a nearly isothermal state (see bottom 
row of figure~\ref{pdf_nofbknosg}) with an average temperature corresponding 
to the minimum of the cooling curve (horizontal dashed line in bottom row of
figure~\ref{pdf_nofbknosg}). Furthermore, the scaling for the 
dispersion of the PDF given by Padoan, Nordlund \& Jones (1997) continued 
to hold. In fact, rather than fit log--normal functions to our density PDFs, 
we measured the average of the logarithm of the gas density, 
$<{\mathrm {log}}_{10} \rho>$, and 
the $M_{\mathrm {rms}}$ of the gas at different
time instances and then overplotted Padoan, Nordlund \& Jones' (1997) 
prediction for the log--normal distribution.
For the runs where we formed stars (without self--gravity or feedback)
in addition to having radiative cooling (runs B1 and C1), 
the gas density PDF continued to have the same behavior:
the $M_{\mathrm {rms}}$ of the system progressively declined with time, while 
the density PDF remained consistent with a log--normal distribution 
(fig.~\ref{pdf_nofbknosgstars}).

\begin{figure*}
\centerline{\psfig{file=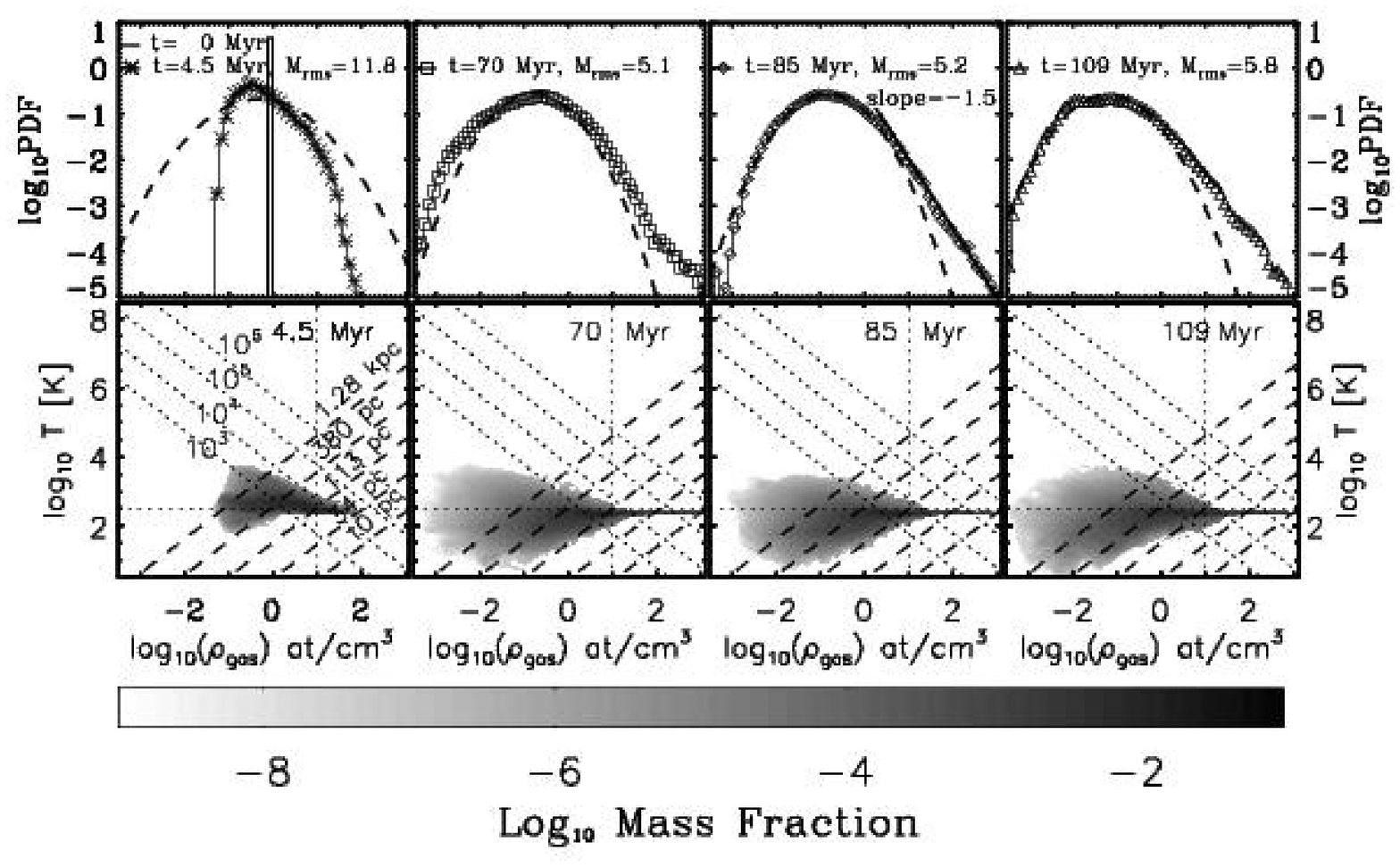,width=.7\hsize,angle=270}}
 \caption{Time evolution of the density PDF (top row) and phase diagrams
(bottom row) for run B2 (128$^3$ run with star formation and self-gravity but
no feedback). The thick dashed line overplotted on the measured
PDFs (symbols) is the log--normal PDF predicted by Padoan, Nordlund \& Jones (1997).
The solid line with a slope of -1.5 plotted at ${\mathrm t = 85}$ Myr
is a fit to the high density end of the PDF. In the phase diagrams, the 
dotted vertical (horizontal) line marks the 
critical density, $\rho_{\mathrm {crit}}$,  (temperature, $T_{\mathrm {crit}}$)
for star formation. Dotted diagonal lines mark lines
of constant pressure, and are labeled for the ${\mathrm t} = 0$ Myr frame:
${\mathrm P} = 10^{6}$, $10^{5}$, $10^{4}$,
and $10^{3}\;{\mathrm k}_{\mathrm B} \;{\mathrm {cm}}^{-3}\; {\mathrm K}$}
\label{pdf_nofbksg}
\end{figure*}
The runs which showed the first departure from log--normal density
PDFs were the runs which included self-gravity (runs B2 and C2)
but still no feedback (figure~\ref{pdf_nofbksg}). Repeating the exercise 
of measuring the average of the logarithm of the gas density, 
$<{\mathrm {log}}_{10} \rho>$, and 
the $M_{\mathrm {rms}}$ of the gas at different times, we found
two differences: (a) the $M_{\mathrm {rms}}$ initially declined 
but then stabilized
at a value higher than that seen in the runs without self--gravity, and
(b) the log--normal PDF predicted by Padoan, Nordlund \& Jones (1997)
consistently underpredicted the distribution at high gas density.
A power--law fit the high density tail well. In one-dimensional simulations
of Burgers flows, i.e. infinitely compressible flows, power--laws
were also found to be good fits to the density PDFs 
(Gotoh \& Kraichnan 1993). We therefore interpret the power--law behavior
for the run with self--gravity, as reflecting the added possibility of
the gas, once it has a high density, to compress to even higher density,
reminiscent of the behavior in Burgers flows.  Klessen (2000) also
explored the form of the density PDF for the cases of decaying and driven
self-gravitating turbulence. Although he found a departure from log-normal 
at high densities, the departure could not be characterized by a power law.

\begin{figure}
\centerline{\psfig{file=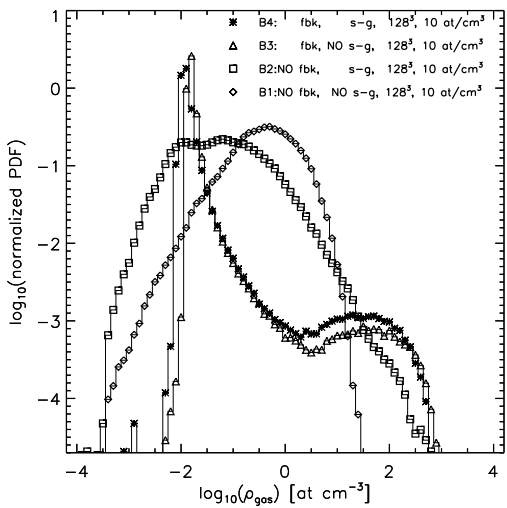,width=.9\hsize}}
 \caption{Comparison of the PDFs at 110 Myrs for runs with different physics.}
\label{pdfcomp}
\end{figure}
\begin{figure}
\centerline{\psfig{file=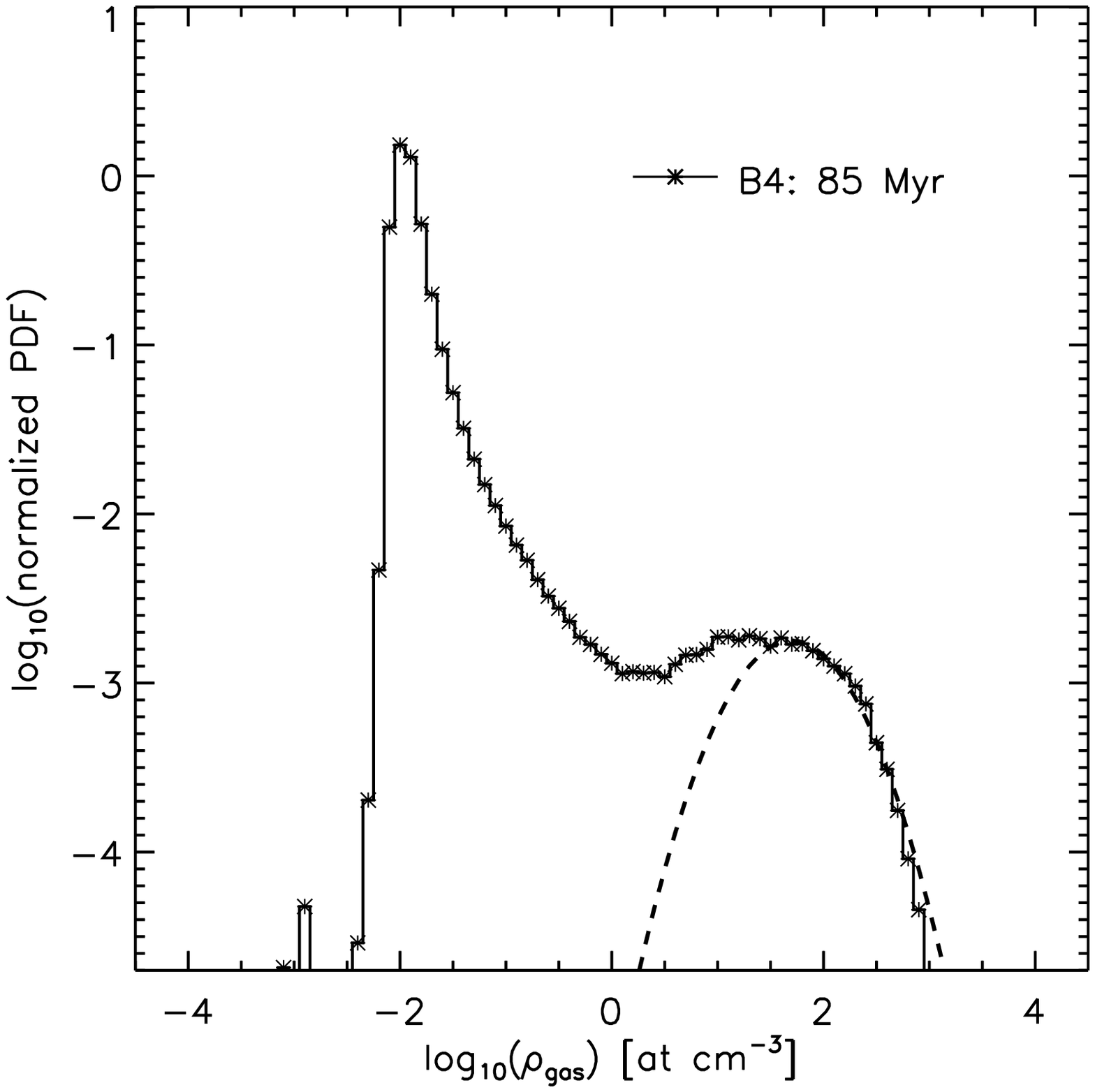,width=.95\hsize}}
 \caption{Log--normal fit to high density end of the PDF for run B4 at
${\mathrm {t = 85} \,\mathrm{Myr}}$. The scaling we use for the fit is a 
log--normal with average density of $10^{1.7}$ atoms/cm$^3$ and with a 
dispersion of $\sim 10^{1.22}$ atoms/cm$^3$.}
\label{pdf_b4_xlnfit}
\end{figure}

When we add feedback to the list of simulated processes,
either with self--gravity (runs B4 and C4) or without (runs B3 and C3),
the density PDF becomes markedly bimodal (figure~\ref{pdfcomp}), 
illustrating that the majority of the simulation volume is occupied 
by low density gas. A bimodal
density distribution is also a sign of a thermal instability 
(V\'{a}zquez-Semadeni, Gazol \& Scalo (2000)) the
consequences of which we will discuss in a future paper 
(Slyz, Devriendt, Bryan \& Silk, {\em in preparation}). For the
runs with self--gravity, the high density power--law tail disappears.
Perhaps it can be argued that the high density part of the density PDF 
may be fit with a log--normal distribution (figure~\ref{pdf_b4_xlnfit}). 
The exercise of overplotting the log--normal given by 
Padoan, Nordlund \& Jones (1997) is not possible because the  
$M_{\mathrm {rms}}$ measured for the entire simulation box does not
correspond to the $M_{\mathrm {rms}}$ of the high density gas for
which the log--normal function may be a good description.
Hence we can only {\em fit} log--normals to the high density gas,
similar to what others, e.g. Wada \& Norman (2001), Kravstov (2003),
do in their global simulations of the ISM. 

The interest of describing the density structure of the ISM
with a single function, such as the log--normal, lies in finding 
a link between the gas density averaged over kiloparsec sized regions
and the high density regions which might form stars.  
This is precisely the link required for an explanation of the Schmidt law. 
Rewriting the Schmidt law in a form where the star formation rate is
equal to some constants multiplied by 
the fraction of gas in high density regions and by the gas density averaged 
over large scales (his equation 7), Elmegreen (2002) emphasized that star 
formation rates depend on the geometry of the density field, i.e. the PDF. 
If the shape of the density PDF is universal, 
then the fraction of gas in high density regions is known. 
Consequently, if the high density 
regions are also self--gravitating, then the fraction of gas available for
star formation is also known. Admittedly, the density PDF contains no
spatial information, hence there is no reason for which the
high density regions should find themselves to be spatially contiguous, 
so that they comprise regions of mass greater than the Jeans mass.  
In fact, figure~\ref{pdf_nofbksg} clearly shows that at least some of the
dense gas regions are not contiguous because if they were they would
simply not persist as all the gas would be converted to stars on a
dynamical timescale since these regions are well above $\rho_{\mathrm {crit}}$
and cold. We therefore have to identify these regions with divergent gas flows.

A two-dimensional study of the ISM in a galactic disk by Wada \& Norman (2001) 
has claimed that the log--normal distribution is a robust description of
the ISM density distribution over many orders of magnitude in density, 
regardless of the simulated physics. More specifically, in their simulations
the presence of stellar feedback does not change the shape of the 
PDF but increases the dispersion of the lognormal. 
In three-dimensional simulations
of a high-redshift galaxy performed in a cosmological context, Kravtsov (2003) 
finds a density distribution similar to Wada \& Norman's (2001). Its
shape at every redshift epoch has a flat region at
$\rho_{\mathrm{gas}} \leq$ 1 -- 10 $\mathrm{M}_{\odot}$ $\mathrm{pc}^{-3}$ and
a power law distribution at high densities. He claims that the log--normal
distribution is a fair description of the high density tail of the PDF and
agrees with Wada \& Norman (2001) on the insensitivity of the distribution
to feedback, except at the low density end, where the simulation with feedback
produces more low density gas. 
As figure~\ref{pdfcomp} shows, our less realistic study of star formation 
occuring in a periodic box without the global gravitational galactic 
potential or the shear instabilities present in a self--gravitating
rotating disk, appears to be more sensitive to the input physics. 
Only the runs which include stellar feedback are nearly equivalent,
regardless of whether there is self-gravity. When log--normals are
overplotted for the runs without feedback, the position of the maximum
of the log--normal is shifted to lower densities by more than one 
order of magnitude from the position of the maximum of the log--normal 
fit to the high density part of the PDF for the runs with feedback. 
Indeed the densities in certain cells for the run with only self--gravity 
reach the same high values as the runs with feedback, but a much smaller 
fraction of the simulation volume has these high densities.
Another blatant difference between the PDFs we find in our runs with feedback
and the PDFs found by Wada \& Norman (2001) and  Kravtsov (2003)
is that their runs do not show as high a peak at low densities.   
The smaller quantity of low density gas in their simulations is likely due 
to the much lower supernovae rates in Wada \& Norman's (2001) simulations 
(0.01 SN/yr as compared to 0.5 -- 4 SN/yr in our simulations) and
in Kravtsov's (2003) case, to the 
more realistic boundary conditions, which allow tenuous, hot gas to escape
the disk.

\subsection{Energy Spectra}
\begin{figure*}
\centerline{\psfig{file=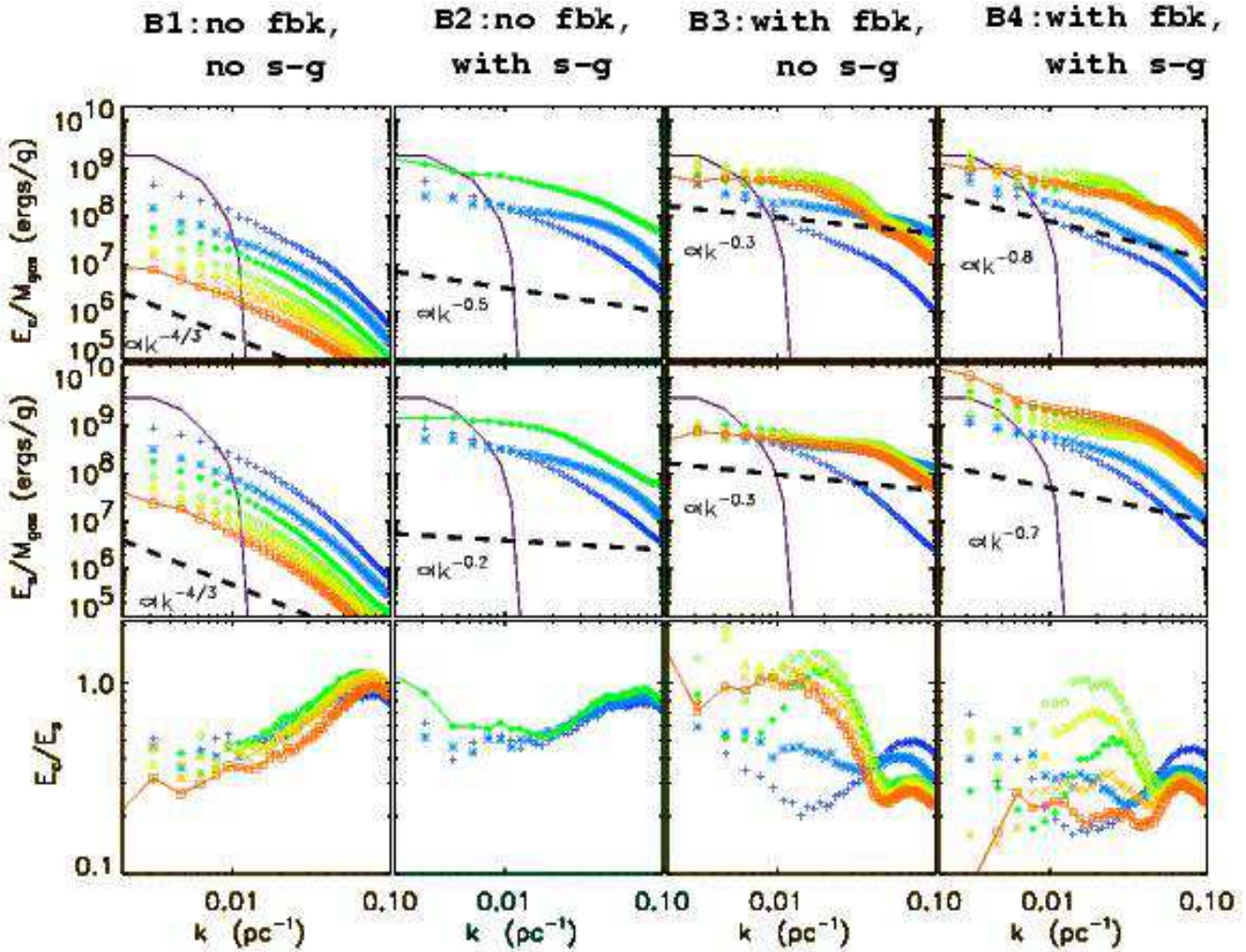,width=\hsize}}
 \caption{Time evolution of the compressible ($E_c$) and solenoidal
($E_s$) components of the energy spectra for runs B1, B2, B3 and B4. Symbols
denote energy spectra at time intervals separated by 30 Myrs. Solid line
represents time $t=0$ Myr. Plus signs: 30 Myr, asterices: 60 Myr, filled
diamonds: 90 Myr, open diamonds: 120 Myr, open triangles: 150 Myr, 
crosses: 180 Myr, open squares: 210 Myr. We also draw a solid line
through the symbols when they represent the final timestep that we
are displaying. Thick dashed lines indicate power laws with slopes
similar to that of the last curve shown.}
\label{energy_spectra}
\end{figure*}
Energy spectra of the ISM carry complementary
information to that given by a study of its density structure. With the 
density PDFs, we confirmed that in many cases there exists a 
clear relationship between the density contrast achieved and the 
$\mathrm{M}_{\mathrm{rms}}$ of a system 
(i.e. $\sigma_{\mathrm{linear}} \sim {\mathrm{M}}_{\mathrm{rms}}$). But the 
${\mathrm{M}}_{\mathrm {rms}}$ of a system is 
only a global measurement of its kinetic energy content.
With measurements of the kinetic energy spectra, we expect to learn how 
the energy is distributed on different spatial scales and how the different 
physical processes we considered influence the time evolution of this 
distribution.

The Kolmogorov theory of incompressible, subsonic turbulence predicts
that energy fed on large scales progressively cascades to smaller scales
until it is dissipated by molecular viscosity on the smallest scales in vortex
rings. The transfer of energy is a local process and 
the spectra of the velocity
field is a power law with $E_k \sim k^{-5/3}$ (Kolmogorov 1941).
With supersonic, compressible turbulence, strong shocks come into play. 
They allow energy to be transferred over widely separated scales and it
is possible that rather than being dissipated in vortex rings, the 
energy is ultimately dissipated in sheets, filaments and cores 
(Boldyrev 2002). Given the analogy between highly supersonic and 
pressureless flows, one might
expect the compressible, supersonic flows to have the same behavior
as Burgers turbulence with power spectra in the inertial regime of the
form, $E_k \sim k^{-2}$ (Burgers 1974, Gotoh \& Kraichnan 1993).  However 
this appears to only be true in one and two dimensions. In three dimensions,
compressible, supersonic flows differ from Burgers flows because they generate 
vorticity (Boldyrev 2002). In three-dimensional simulations of 
compressible, supersonic,
magnetized forced turbulence with Mach number initially $\sim$ 10,
Boldyrev, Nordlund \& Padoan (2002) find energy power spectra in the
inertial range to be  $E_k \sim k^{-1.74}$, i.e. close to the Kolmogorov
value.

As we lack the grid resolution to ascertain if the energy spectra
in our simulations are tending towards power laws, we cannot make any 
credible statements 
about the values of the power law slopes.
Furthermore in incompressible turbulence, the energy spectrum is a power
law in the inertial regime (at $k$ wavenumbers below the energy injection
scale but above the energy dissipation scale). In our simulations
the feedback energy is injected on scales equivalent to the grid resolution,
i.e. the smallest scales, but it can propagate to larger scales depending
on the ISM dynamics. Therefore for the runs with feedback
the inertial regime has a more complicated meaning.
Instead in figure~\ref{energy_spectra} we focus on the time development
of the energy spectra, and the presence of characteristic features.
The standard approach involves dividing the kinetic energy
into two components:  a compressible one for which 
$\nabla \times \; v_{\mathrm {comp}} = 0$, and a solenoidal one with 
$\nabla \cdot \; v_{\mathrm {sol}} = 0$. In words, the compressible component
measures the strength of the shocks in the system, while the solenoidal
component measures the degree of rotation. Typically, the compressible
component is expected to decay faster than the solenoidal component
as the shock energy is transformed into vortical eddy motions. 

Because we remove gas from the system to form stars, the kinetic
energy whose spectra we measure, is rather a specific kinetic energy,
i.e. we divide the instantaneous total kinetic energy by the 
total gas mass present at that moment. In all our runs, the
kinetic energy which is initially imprinted only on large scales
quickly (within $\sim$ 30 Myr) redistributes itself to smaller
scales as well.  Following this redistribution, for the run with
neither self--gravity nor feedback (run B1), the compressible and solenoidal
components of the energy spectra progressively decay all the while
maintaining approximately the same form. The ratio $E_c/E_s$ is always
less than 1, i.e. the compressible component decays faster
than the solenoidal one, but increases towards the dissipative regime. 
In high resolution simulations ($512^{3}$, $1024^{3}$) of decaying 
compressible turbulence with Mach number initially on the order of 1 (an 
order of magnitude lower than the initial Mach number in our simulations), 
Porter, Woodward \& Pouquet (1998) find a similar result with 
$E_c/E_s \sim 0.1$.
In contrast to these runs in which the kinetic energy decays, the runs
with self--gravity (run B2) and/or feedback (runs B3, B4), 
show energy spectra which
climb to higher amplitudes with time and have shallower slopes than the 
decaying run (B1). Furthermore in plots of the ratios of the compressible to 
solenoidal components, between 90 and 150 Myrs the runs with
feedback show a peak at $\sim$ 65 pc consistent with what one
would predict for the characteristic lengthscale for a simulation
with supernovae expanding into a medium with ambient pressure of
$P = 10^{6} cm^{-3} \; K \; k_{B}$. More explicitly, 
ignoring adiabatic and radiative losses, 
a supernovae with $10^{51}$ ergs of energy will be halted by an ambient 
medium at this pressure when it has expanded to a radius, 
$r \sim (E/P)^{1/3} \sim$ 65 pc. This signature in $E_c/E_s$ for
the run with feedback points to a way to understand SFRs, which
we explore below.

\section{Numerical versus analytical star formation rates}
\label{comparesilk}
An alternative to searching for a generic density PDF
as an explanation for star formation rates, is to
consider arguments concerning the competition between the expansion
of supernovae remnants and the pressure which halts them.
In this vain, Silk (1997, 2001) developed porosity models
of a regulated ISM. Introduced by Cox and Smith (1974), 
porosity, ${\mathrm Q}$, is proportional to the product of the
supernovae rate per unit volume and the maximum extent of the 
4--volume of the supernovae remnants. In other terms, the porosity
measures the fraction of hot gas, ${\mathrm f_h}$, 
in the ISM through the relation  ${\mathrm Q} = 
-{\mathrm {ln}}\,(1 - {\mathrm f_h})$.
Silk reasoned that since the supernovae production rate is proportional to the
star formation rate (SFR), and the maximum extent of
a supernovae remnant is limited by the ambient pressure, the
following expression arises:
\begin{equation}
{\mathrm Q} = {\mathrm {SFR}} \; {\mathrm G}^{-1/2}\; \rho_{\mathrm {gas}}^{-3/2} \; (\sigma_{\mathrm {gas}}/\sigma_{\mathrm f})^{-2.72}
\label{sfrequation}
\end{equation}
where $\rho_{\mathrm {gas}}$ is the gas density, $\sigma_{\mathrm {gas}}$ is
the gas velocity dispersion, and $\sigma_{\mathrm f}$ is a fiducial
velocity dispersion that is proportional to 
$E_{\mathrm {SN}}^{1.27}\,m_{\mathrm {SN}}^{-1}\,\zeta_{\mathrm g}^{-0.2}$.
Here $E_{\mathrm {SN}}$ is the energy of a single supernova,
$\zeta_{\mathrm g}$ is the metallicity relative to solar of the ambient gas,
and $m_{\mathrm {SN}}$ is the mean mass in newly formed stars required to produce
a supernovae. For $E_{\mathrm {SN}}$ = $10^{51}$ erg, $\zeta_{\mathrm g} = 1$,
and $m_{\mathrm {SN}} = 250 M_{\odot}$ i.e. the case where we assume
only the occurrence of type II supernovae with a Miller--Scalo IMF,
the fiducial velocity dispersion is $\sim$ 22 km ${\mathrm s}^{-1}$.

Our simulations with feedback provided a laboratory to test 
this analytic description of the SFR. For the purpose of computing
the porosity of the medium, we measured the fraction of hot gas in our
volume, defining hot to be gas with temperature T $\geq$ 4 $\times 10^{6}$ K. 
For $\rho_{\mathrm {gas}}$
in equation~\ref{sfrequation} we took the average gas density in
our simulation volume, and for $\sigma_{\mathrm {gas}}$ we took the average
mass--weighted velocity dispersion of the gas. We kept the value for 
$\sigma_{\mathrm f}$ at 22 km ${\mathrm s}^{-1}$. Given these values as
functions of time, we plotted as dotted and dashed lines the expectation
from eq.~\ref{sfrequation} for the SFRs in figure~\ref{sfrs}.
Computing the actual star formation rates in the box by defining the
mass of newly formed stars to be the mass of stars formed in the past
3 Myr, we overplotted the results as symbols in the same figure. Astonishingly,
the analytic values match the measured rates to better than
a factor 2. 

Given the simplifications in the derivation of the analytic model,
there was no {\em a priori} reason for the fit to be a good
description of the star formation rate in an inhomogenous, non--stationary
model of the ISM. For example,  
Silk takes the expression for the 4--volume of the SNR remnant in its
cooling phase from Cioffi, Mckee \& Bertschinger
(1988).  They derive it under the assumptions that the supernovae expands
in a spherical manner, the ISM is homogenous and uniform 
(i.e. no density gradients), there is no dust cooling or thermal conduction,  
and the ambient ISM pressure is negligible until the
last stage of supernovae evolution when the remnant merges with the ambient
ISM. In contrast, we find that at least in the initial stages of our 
simulations,
the supernovae remnants are highly non--spherical, the ISM is inhomogeneous
with ubiquitous density gradients and the ambient ISM gas pressure is
highly non-negligible ($P = 10^{5}-10^{6} cm^{-3} \; K \; k_{B}$).  
However, as more of the gas turns into
stars, and the hot phase fills the majority of the simulation volume, the
ISM does start to resemble something more in line with the Cioffi et al.
assumptions.

\begin{figure}
\centerline{\psfig{file=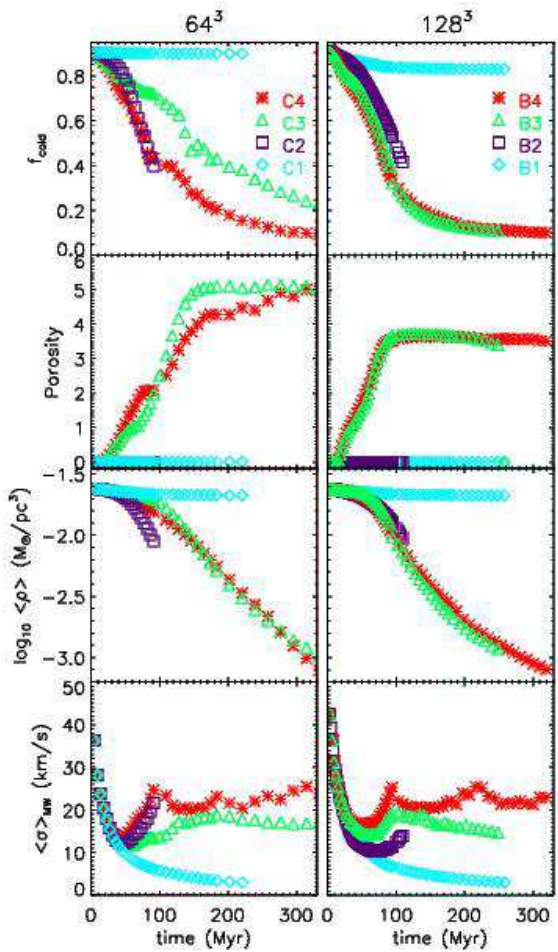,width=1.5\hsize,angle=270}}
 \caption{Plots comparing the time evolution of the cold mass fraction, 
porosity, average gas density, and mass-weighted velocity dispersion for runs
containing different physics, on the $64^3$ and $128^3$ grid.}
\label{massfprhosigma64and128}
\end{figure}
When we examine in figure~\ref{massfprhosigma64and128} the time 
evolution of each of the 
physical quantities entering into the analytic model
for the SFR, we find the following. The runs (B1 and C1) which 
produced the lowest star formation rates have zero porosity
and high fractions of cold gas ($f_{\mathrm{cold}} \sim$ 0.8--0.9), 
but a continuously declining velocity dispersion. The runs reaching 
a peak (runs B2, B3, B4, C2, C3) 
or multiple peaks of high star formation (run C4) all
displayed depleted cold gas fractions after their final star formation
peak, a rise to a maximum in its velocity dispersion at the peak, and 
either zero porosity for the case of the runs with self--gravity but 
no feedback (runs B2 and C2) or a porosity that levels off to a constant
value around the time of the SFR peak (Q $\sim$ 4--5 for the $64^3$ case 
(runs C3 and C4), and Q $\sim$ 4 for the $128^3$ case  (runs B3 and B4)
after the SFR peak). We interpret the behavior in these parameters
as reflecting the importance of a high velocity dispersion for 
generating high SFRs. Indeed in the analytic model for the SFR 
(eq.~\ref{sfrequation}),
the gas velocity dispersion, $\sigma_{\mathrm {gas}}$, plays the most
important role, as it is raised to the highest power in the expression.  
However even with velocity dispersions sustained at high values 
($\sigma_{\mathrm{gas}} \sim$ 20 km/s),
 SFRs will drop if the reserves of cold gas decline. 

\section{Discussion}
\label{discussion}

Given the simplicity of our simulations, we examine their relevance
for representing true star formation processes in real galaxies.
The first issue we address is whether the star formation rates we obtain
are consistent with the Kennicutt relation. In section~\ref{detailedpics}
we scaled the mass in our simulation volume to that of the Milky Way,
finding that our star formation rates and surface densities were consistent
with star formation occuring in the starburst regime.
If we do not scale our SFRs and gas densities to a Milky Way type galaxy but
instead take them at face value we find that our initial 1 atom/cm$^3$ gas
density in a (1.28 kpc)$^3$ volume yields in projection about a 30 M$_\odot$/pc$^2$ column
density which lies at the boundary between Kennicutt's normal disks and 
centers of normal disks (Kennicutt 1998). Transforming our average star formation
rate of 0.2-0.3 M$_\odot$/yr into a star formation rate per unit volume
leads us to an average star formation
rate density of about 0.1 M$_\odot$/yr/kpc$^2$, on the high side but in fair
agreement with Kennicutt's measurements for our computed surface density 
(Kennicutt 1998, figure 6). We note
that Kennicutt's law is a static relation as it
concerns space averaged quantities in local galaxies, and a moment in the
history of these galaxies is bound to exist when their main progenitor
will be entirely gaseous (i.e. with no stars yet formed) and the Kennicutt relation
will break. As our simulations start from an exclusively gaseous medium, we do
not expect our simulation to follow the Kennicutt
relation from the very beginning, but to move towards it as it does.
We nevertheless consider our simulations to be in a starburst mode because 
the duration of the star formation episode is much shorter than that of what
one expects in either a disk or spheroidal 
galaxy. But this is not unusual since we are only modeling a chunk of 
a galaxy and are therefore neglecting effects on larger length and therefore
timescales.

The second issue we address is whether  periodic 
boundary conditions drive the high star formation rates seen in our simulations. 
When hot gas starts to fill the bulk of the simulation volume, because
the boundary conditions trap the hot gas, 
conditions in the simulation may be viewed as a pressure cooker
and the increased pressure may drive higher star formation rates.
In our simulation by the time the
pressure cooker is operative, the SFRs are already at starburst levels as seen
when one
scales the SFRs and gas densities to a Milky Way type galaxy as we do in section~\ref{detailedpics}.  
To be more specific, for the pressure cooker to be operative we have to 
wait $\sim$ 10 Myr for the first supernovae to go off
and then we have to wait for the volume to
become significantly filled by this supernovae generated hot gas for the 
hot gas to be able to
traverse the volume unobstructed by cold, dense gas. According to figure 15,
it takes on the order of 50 Myr for the hot gas filling fraction to be
approximately 50\%, corresponding to a porosity of about 0.7. Hence boundary
effects are not dominant in shaping the star formation rate until after that
time. 
We also point out that the limitations of the boundary conditions should not
obfuscate the point that the manner in which we implement supernovae is a more important
factor leading to the build up of large quantities of hot gas in the medium.  
When we perform simulations in all points identical to those presented in this
paper but with 
supernovae going off instantaneously, as opposed to exploding with a more realistic 10
Myr time delay used in the work presented in this paper, we get extremely low
star formation rates (a few hundred times smaller than those we get in our
simulation here), because the hot gas never fills a significant fraction of
the simulation cube.  In other words, the periodic boundary conditions cannot
dominate the physics of star formation driven by hot gas pressure until the
hot gas has already been generated, and we find that this depends strongly on
the way the
supernovae are implemented. As mentioned in section~\ref{introduction}, we leave the
discussion of this to a future paper.

The limitations of our closed, periodic box, and the
absence of a stratified external gravitational potential certainly keep our simulations far
from being representative of realistic galactic systems.  For example, a credible simulation
of a disk galaxy, would have to be performed in a realistic cosmological
context to capture such effects as tidal encounters and stripping from
neighbours. Excluding these external stellar heating processes as well 
as spiral waves, results in the neglect of processes that
would increase the velocity dispersion of the stars in real
galaxies. Therefore our simulations certainly have a higher 
fraction of cold ISM and cold stars after a gas consumption time which may
prolong and strengthen star formation in our simulations.

We also emphasize that with our crude assumption of a
closed box not only can no material escape the box, affecting star formation
rates once hot gas permeates the simulation volume, but no material can
enter the simulation volume either. It could well be that accretion of cold
material is more relevant for star formation in real disks than either the
external star heating processes missing from the simulations discussed above or
the fact that hot gas cannot leave the simulation volume.
One can argue that perhaps the simulations presented in this paper 
are more representative of what happens in the central
kiloparsec of a spheroidal starburst galaxy. In that case the potential well
might indeed trap a fraction of the hot gas and the pressure cooker
environment which comes into play after high star formation rates occur
in the simulation, if not as drastic as in our simulations might well be fairly
realistic.

\section{Summary and conclusions}
\label{conclusion}
To unravel which global parameters control star formation, we 
have examined star formation occurring in media whose dynamics
are structured by various combinations of physical processes (e.g.
``turbulence'', radiative cooling, self-gravity, feedback from 
supernovae and stellar winds). We sought to understand our models
of the ISM from structural and dynamical perspectives, finding
that in some cases there was a well-defined link between the two.
In particular, measurements of the density PDFs confirmed that
for the simulations without feedback, lognormals were an adequate
description of the structure of the medium, and that the density contrasts 
achieved in the media were directly correlated to their 
${\mathrm{M}}_{\mathrm{rms}}$. Lognormals consistently underpredicted
the high density end of the runs with self-gravity which appeared to
be well-fit by a power law. For the runs with feedback, the
dense gas reached higher densities than those reached by the
runs without feedback implying that in these simulations, feedback
was positive in the sense that it encouraged higher star formation rates.
However the PDF for the runs with feedback had a distinctly bimodal shape
with the majority of the volume filled by low density gas. In summary, we
did not find a universal PDF. Most markedly, runs with feedback had a different
PDF from the runs without feedback, although arguably, the high density
end might be fit by a lognormal.

Measurements of the energy spectra in our simulations were consistent
with the information provided by the density PDFs. Self--gravity alone
was sufficient to sustain the kinetic energy of the medium, and hence
maintain the high density contrast we observed in the PDFs. Feedback
also succeeded in keeping high quantities of kinetic energy in the media
and inspection of ratios of compressible to solenoidal energy revealed
that supernovae were pumping energy into the system at a characteristic
scale consistent with the ambient pressure in the hot, low density
component of the medium.

For the runs with feedback, comparing Silk's (2001) star formation model to the measured
values of the SFRs in our simulations, revealed a good match that
led us to inspect the parameters involved in Silk's prescription.
They showed clearly that the SFR depends strongly on the
underlying velocity field which we saw could be energized 
by self--gravity and/or feedback to produce high density contrasts. Without
a means to create these high densities, star formation rates decline
even in the presence of a large reservoir of cold gas.


In light of the issues neglected in our simulations, we
stress that the simplifying assumptions made in this paper facilitated our
choice to start from as strong as possible a local physical basis as possible
before trying to tackle star formation in a more global context. As such
we neglect numerous physical processes which may invalidate partially or completely our current
results, but this remains to be addressed in future work. 
Nevertheless we hope that the present work sheds some light on the local
physics that should be included in future realistic simulations of star
formation.



\section*{Acknowledgments}
The authors thank Fabian Heitsch for a careful reading of the manuscript.
A. Slyz acknowledges the support of a Fellowship from the UK
Astrophysical Fluids Facility (UKAFF) where some of the computations reported
here were performed.


\end{document}